\documentclass[referee,sn-nature]{sn-jnl}

\usepackage{graphicx}%
\usepackage{multirow}%
\usepackage{amsmath,amssymb,amsfonts}%
\usepackage{amsthm}%
\usepackage{mathrsfs}%
\usepackage[title]{appendix}%
\usepackage{xcolor}%
\usepackage{textcomp}%
\usepackage{manyfoot}%
\usepackage{booktabs}%
\usepackage{algorithm}%
\usepackage{algorithmicx}%
\usepackage{algpseudocode}%
\usepackage{listings}%
\usepackage{geometry}%
\theoremstyle{thmstyleone}%
\newtheorem{theorem}{Theorem}
\theoremstyle{thmstyletwo}%

\theoremstyle{thmstylethree}%

\begin{document}

\title[Three-dimensional Moir{\'e} crystallography]{Three-dimensional Moir{\'e} crystallography}


\author*{\fnm{Ilya} \sur{Popov}}\email{Ilya.Popov1@nottingham.ac.uk}

\author*{\fnm{Elena} \sur{Besley}}\email{Elena.Besley@nottingham.ac.uk}

\affil{\orgdiv{School of Chemistry}, \orgname{University of Nottingham}, \orgaddress{\street{University Park}, \city{Nottingham}, \postcode{NG7 2RD}, \country{United Kingdom}}}


\abstract{\textbf{Moir{\'e} materials, typically confined to stacking atomically thin, two - dimensional (2D) layers such as graphene or transition metal dichalcogenides, have transformed our understanding of strongly correlated and topological quantum phenomena. The lattice mismatch and relative twist angle between 2D layers have shown to result in Moir{\'e} patterns associated with widely tunable electronic properties, ranging from Mott and Chern insulators to semi- and super-conductors. Extended to three-dimensional (3D) structures, Moir{\'e} materials unlock an entirely new crystallographic space defined by the elements of the 3D rotation group and translational symmetry of the constituent lattices. 3D Moir{\'e} crystals exhibit fascinating novel properties, often not found in the individual components, yet the general construction principles of 3D Moir{\'e} crystals remain largely unknown. Here we establish fundamental mathematical principles of 3D Moir{\'e} crystallography and propose a general method of 3D Moir{\'e} crystal construction using Clifford algebras over the field of rational numbers. We illustrate several examples of 3D Moir{\'e} structures representing realistic chemical frameworks and highlight their potential applications in condensed matter physics and solid-state chemistry.}}


\maketitle

\noindent Moir{\'e} physics, based on superposing two or more two-dimensional (2D) crystals with a relative twist between the layers, have created many interesting nanomaterials and thin films with a wide range of unique optical, magnetic, and electronic properties \cite{Andrei_2021,He_2021,Mak_2022}. The lattice mismatch, required for the formation of Moir{\'e} patterns, appears naturally when the parameters defining the unit cell of the constituent lattices are not commensurate ($\it i.e.$ not in proportion), or it can be engineered by rotation and displacement of the constituent lattices. The latter approach has evolved into  a separate active research field in condensed matter physics called twistronics \cite{Carr_2017,Hennighausen_2021}. Starting from experiments on twisted graphene layers \cite{Andrei_2020,Li_2009,Cao_2018,Luican_2011}, twistronics has been recently extended to other materials including hexagonal boron nitride \cite{Cho_2024,Kim_2023}, transition metal dichalcogenides \cite{Tao_2024}, and layered cuprates \cite{Zhao_2023}.

In mathematics, 2D Moir{\'e} patterns are constructed from a given prototype lattice, $L$, by applying an in-plane rotation, $r$, and overlapping the twisted layer, $rL$, with the initial one. Rotation $r$ belongs to the group $SO_{2}\left(\mathbb{R}\right)$ which represents the set of all rotations in a 2D Euclidean space, and $L\cong\mathbb{Z}^{2}$ is a two-dimensional Bravais lattice. If the lattices $L$ and $rL$ are commensurate with respect to translations in two dimensions, then their overlap forms a periodic Moir{\'e} pattern $L\cup rL$. In some cases, an additional in-plane displacement can be applied to the layer $rL$. Overlapping of commensurate $\mathbb{Z}^{2}$ infinite lattices is always possible in three-dimensional (3D) space without any steric hindrance by embedding them into parallel planes. In principle, the construction of Moir{\'e} crystals can be generalised to three dimensions by considering lattices $L\cong\mathbb{Z}^{3}$ and rotations belonging to the $SO_{3}\left(\mathbb{R}\right)$ group. 

The initial steps towards such generalisation have been made in recent works \cite{Wang_2024,Gao_2025}, where Moir{\'e} physics has been extended to three dimensions by considering crystals produced by the overlap of two simple cubic lattices twisted with respect to one another. In ref.~\cite{Wang_2024}, it was suggested that the formed 3D Moir{\'e} structures can be used as theoretical models of ultra-cold atomic gases suitable for optics applications. However, to enable an entirely new branch of crystallography dedicated to studying periodic 3D Moir{\'e} structures, several important questions of both mathematical and chemical nature need to be addressed. The mathematical problem can be formulated in the following way. What are the conditions that an arbitrary prototype lattice $L\cong\mathbb{Z}^{3}$ must satisfy in order for rotations $r$ to exist such that the Moir{\'e} pattern $L\cup rL$ is periodic with respect to 3D translations ($\it i.e.$ $L\cup rL\cong\mathbb{Z}^{3}$)? And if such rotations exist for a given lattice $L$, can they be fully parametrised and classified? Wang $\it et~ al.$ \cite{Wang_2024} showed that for a particular case of  simple cubic lattice the allowed rotations generating Moir{\'e} crystals belong to $SO_{3}(\mathbb{Q})$ group, which can be parametrised by a set of five integer numbers. In crystallography, however, lattices can belong to seven different systems of various symmetries \cite{Crystal_2006}, with the cubic lattice being a single and simplest example. 

In this work, we present a general fundamental solution to the mathematical problem of constructing 3D periodic Moir{\'e} crystals which includes all possible lattice structures. We formulate the necessary conditions for the existence of periodic Moir{\'e} patterns for any arbitrary $\mathbb{Z}^{3}$ lattice, and, using Clifford algebras over the rational field, we give a complete parametrisation of the manifolds of rotations generating these patterns. This allows us to propose a complete crystallographic classification of 3D Moir{\'e} crystals. In relation to chemistry and physics of 3D Moir{\'e} crystallography, we further address a question of whether 3D crystals created in this way may represent a realistic stable or metastable  phase of solid-state matter. Although constructed in a way similar to 2D Moir{\'e} materials, 3D structures have different embedding into 3D space and exhibit distinct bonding patterns. As a result, their potential applications will go well beyond nanomaterials and devices developed by the conventional twistronics. To illustrate this, we generate various examples of novel 3D Moir{\'e} crystals representing chemically meaningful frameworks and analyse their structure and topology from crystallographic point of view. This work lays the principal foundations of 3D Moir{\'e} crystallography.

\section*{General construction of 3D Moir{\'e} crystals}\label{sec:general}

\noindent In a Cartesian coordinate frame, for a 3D lattice $L\cong\mathbb{Z}^{3}$ with non-coplanar basis vectors $\boldsymbol{\mathbf{u}}_{1}$, $\boldsymbol{\mathbf{u}}_{2}$, and $\boldsymbol{\mathbf{u}}_{3}\in\mathbb{R}^{3}$, the unit vectors are given by a matrix

\begin{equation}
u=\left(\begin{array}{ccc}
u_{1x} & u_{2x} & u_{3x}\\
u_{1y} & u_{2y} & u_{3y}\\
u_{1z} & u_{2z} & u_{3z}
\end{array}\right)\label{eq:umat}
\end{equation}
\\
\noindent with real elements and $\det u\neq0$. t In crystallography, a crystal lattice is defined by a set of six  parameters $\left(a,b,c,\alpha,\beta,\gamma\right)$ describing the unit cell. They can be used to express the Gram matrix as

\begin{equation}
g=\left(\begin{array}{ccc}
a^{2} & ab\cos\gamma & ac\cos\beta\\
ab\cos\gamma & b^{2} & bc\cos\alpha\\
ac\cos\beta & bc\cos\alpha & c^{2}
\end{array}\right),\label{eq:gram}
\end{equation}
\noindent where $g$ is symmetric and $\det g>0$. Representation (\ref{eq:gram}) of the Gram matrix is routinely used in crystallographic structure classification to associate any given lattice $L$ with one of the seven known crystal systems \cite{Crystal_2006}. The Gram matrix (\ref{eq:gram}), therefore, fully defines lattice $L$, and it plays a key role in the construction of 3D Moir{\'e} crystals from the prototype lattice $L$. The unit cell of $L$ contains atoms whose positions are specified by the fractional coordinates $\left\{ f_{ij}\right\} _{i=1\ldots N}^{j=1\ldots3}$, where $N$ is the number of atoms in the unit cell. The unit cell parameters and a set of the fractional coordinates of atoms determine the space group of the crystal associated with lattice $L$ \cite{Crystal_2006}.

\begin{figure}[ht]
\centering
\includegraphics[width=0.95\textwidth]{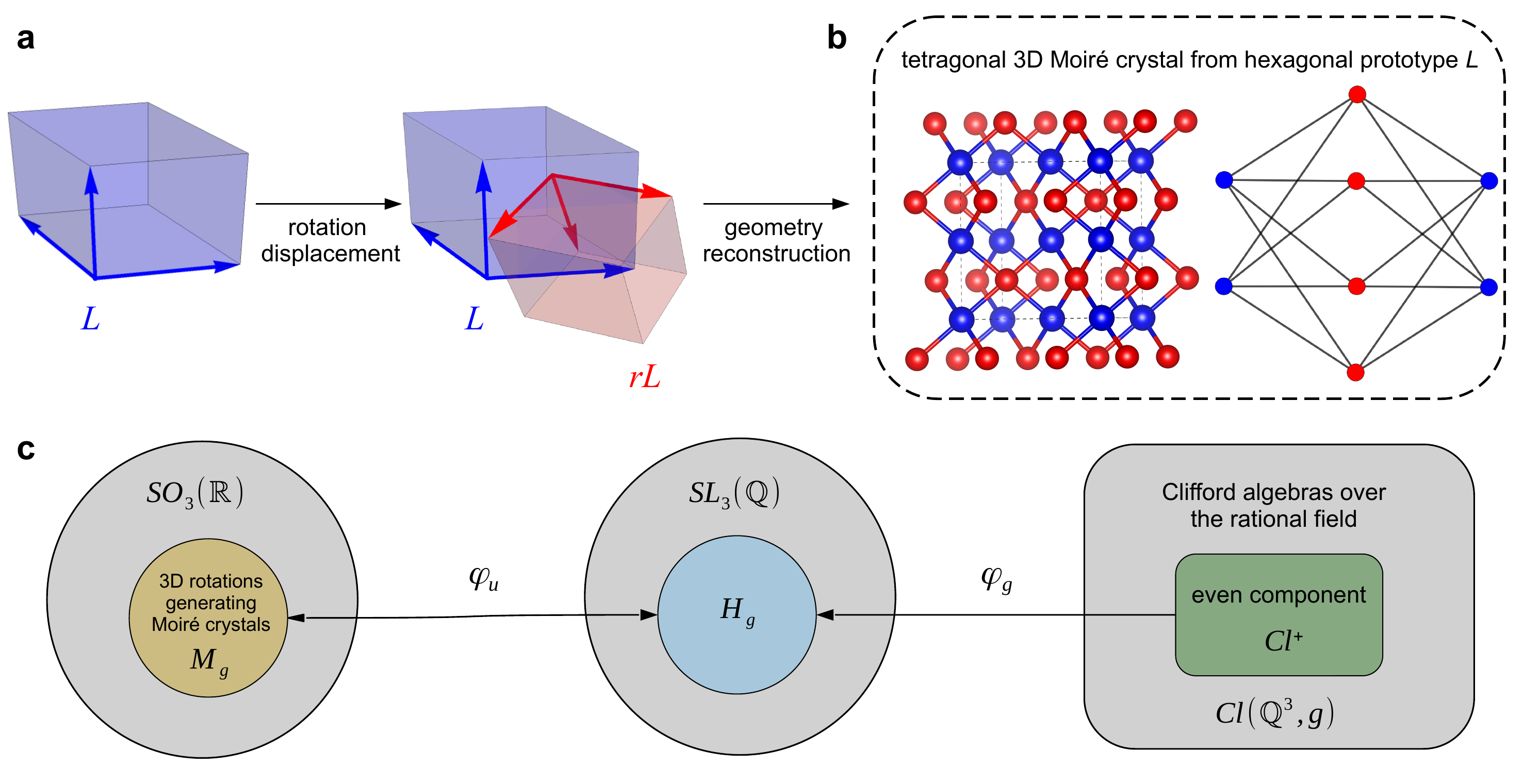}
\caption{\textbf{Pictorial and mathematical construction of 3D  Moir{\'e} crystals.}
		(\textbf{a}) A superposition of the prototype lattice $L$ (blue) and lattice $rL$ (red) formed by the rotation and subsequent displacement of $L$ can create a 3D Moir{\'e} superlattice.  (\textbf{b}) The primitive unit cell of the prototype lattice $L$ contains one atom, whilst reconstruction of the atomic positions and bonds in the corresponding 3D Moir{\'e} superlattice gives a crystalline structure with eight atoms in the unit cell (left). The resulting 3D Moir{\'e} crystal is a fully connected bipartite chemical network with four-coordinated square-planar sites. The topology of this crystal is characterised by the fundamental graph, $X_0$ (right).  (\textbf{c}) Schematic representation of the mathematical objects used to derive 3D Moir{\'e} crystals and morphisms between them. The group of rotations $M_{g}$ that generates all possible Moir{\'e} lattices for a given Gram matrix $g$ is mapped to the $SL_{3} (\mathbb{Q})$ group and evaluated using Clifford algebras defined over the field of rational numbers.}
\label{fig:scheme}
\end{figure}

Let us consider a rotation $r\in SO_{3}\left(\mathbb{R}\right)$ transforming the lattice $L$ into lattice $rL$ with the basis vectors $\mathbf{u}'_{i}=r\boldsymbol{\mathbf{u}}_{i}$. Overlapping two lattices, $L$ (blue lattice in Fig.~\ref{fig:scheme}a) and $rL$ (red lattice in Fig.~\ref{fig:scheme}a), produce a 3D Moir{\'e} crystal if and only if they are periodically commensurate with each other. This requires the existence of a rational matrix $h\in SL_{3}\left(\mathbb{Q}\right)$ satisfying the following equation

\begin{equation}
\left(\begin{array}{c}
\mathbf{u}'_{1}\\
\mathbf{u}'_{2}\\
\mathbf{u}'_{3}
\end{array}\right)=h^{t}\left(\begin{array}{c}
\mathbf{u}{}_{1}\\
\mathbf{u}{}_{2}\\
\mathbf{u}{}_{3}
\end{array}\right).
\end{equation}
\\
\noindent In matrix notations, the relationship between the rotation matrix $r$ and rational matrix $h$ has the form $ru=uh$. Our goal, therefore, is to find rotations, $M_{g}$, for which the matrices $u^{-1}ru$ are rational (note, that  $r$ and $u$ are not necessarily rational), where $M$ stands for ``Moir{\'e}'' and index $g$ emphasises that this set depends on the choice of the lattice. 

 We first discuss the general principles of constructing 3D Moir{\'e} crystals. Suppose we found a rotation matrix $r$ such that $h=u^{-1}ru$ is rational. The elements of matrix $h$ can be expressed as $h_{ij}=m_{ij}/n_{ij}$, where $m_{ij}$ and $n_{ij}$ are co-prime integers. For a set of three integer numbers $l_{i}=\mathrm{lcm}\left(n_{1i},n_{2i},n_{3i}\right)$ and a matrix with the following elements $k_{ij}=l_{i}m_{ji}/n_{ji}$, the unit cell of a 3D Moir{\'e} lattice $L\cup rL$ is spanned by vectors $l_{i}\mathbf{u}'_{i}=\sum_{j}k_{ij}\boldsymbol{\mathbf{u}}_{j}$ and it contains two types of atoms originating from the constituent crystals $L$ and $rL$ (shown in blue and red in Fig.~\ref{fig:scheme}b). The fractional coordinates of the $i$-th atom belonging to $L$ in the constructed Moir{\'e} crystal can be determined as $\bar{f}_{i}=k^{-t}f_{i}$, while the coordinates of the $j$-th atom belonging to $rL$ is determined as $\bar{f}_{j}=lf_{j}$, where $l=\mathrm{diag}\left(l_{1},l_{2},l_{3}\right)$. In both cases, the indices $i$ and $j$ run through all atoms of crystals $L$ and $rL$ for which the resulting coordinates $\bar{f}$ lie in the interval $\left[0,1\right)$. If positions of any two atoms in the constructed unit cell coincide, they are replaced by a single atom. We note, that a set of unit cell vectors $\left\{ l_{i}\mathbf{u}'_{i}\right\} $ might not reflect overall symmetry of the Moir{\'e} crystal. Due to that, the resulting unit cell must be transformed to the Niggli reduced cell~\cite{Niggli,Santoro_1970} to ensure correct assignment of the lattice system.

When constructing 3D Moir{\'e} crystals, we might use not only rotation but also displacement of the lattices $L$ and $rL$ relative to each other. A displacement vector $\mathbf{d}$ can always be expanded in the basis of the unit cell vectors $\left\{ l_{i}\mathbf{u}'_{i}\right\} $. Therefore, vector $\mathbf{d}$ shifts the fractional coordinates of atoms belonging to $rL$ without affecting periodicity of the Moir{\'e} lattice. This fact allows us to choose vector $\mathbf{d}$ arbitrarily without concerns about the commensurability of lattices $L$ and $rL$.

To ensure that a Moir{\'e} crystal constructed mathematically represents a meaningful chemical structure, we introduce bonds between atoms in the following way. If the minimal distance between atoms in a crystal is denoted as $D$, two atoms are considered to be bonded if the distance between them does not exceed $sD$, where parameter $s>1$ determines a range of bond lengths allowed in a given crystal (in this work, we take $s=1.2$). The atoms in the Moir{\'e} crystal together with the bonds form a chemical network characterised by an infinite graph $X$ \cite{Sunada_2013}. Its finite fundamental graph $X_{0}$ is defined as a quotient of $X$ with respect to the translational group of the lattice, and it determines the topology of the 3D network \cite{Sunada_2013}. To demonstrate the relevance of fundamental graphs to our analysis we give an example of tetragonal 3D Moir{\'e} crystal together with its fundamental graph $X_{0}$ (Fig.~\ref{fig:scheme}b). In this structure, every site belonging to the prototype lattice $L$ is bonded to four sites of the lattice $rL$ and vice versa. Atoms and bonds form a connected infinite network, that can not be partitioned into subsystems without breaking the bonds. This bonding pattern is vividly reflected in the topology of the fundamental graph $X_{0}$, that can be analysed by numerical methods of the graph theory. Such analysis is particularly useful for chemical networks with large unit cells and complicated connectivity patterns which we shall encounter further.

One of the most important characteristics of $X_{0}$ is the number of components. Generally, we distinguish three possible cases depending on the number of components in the fundamental graph $X_{0}$: (i) a single component graph means that the constructed Moir{\'e} crystal represents a realistic 3D chemical framework; (ii) if the number of components is larger than one but significantly smaller than the number of atoms in the unit cell then the Moir{\'e} crystal corresponds to either a layered solid material or a periodic packing of finite clusters; (iii) if the number of components is comparable to the number of atoms in the unit cell, then the constructed 3D object does not correspond to any realistic solid-state material. Based on this classification, the crystal shown in Fig.~\ref{fig:scheme}b belongs to the first class and represents realistic chemical 3D network. In the next Section, we shall discuss examples of 3D Moir{\'e} crystals of the first and second kind in more detail.

The problem of finding a set of rotations $M_{g}$ that generate commensurate Moir{\'e} patterns for a given Gram matrix $g$ is crucial in the described construction. The structure of $M_{g}$ is discussed in Supplementary Text S1, where we prove that $M_{g}$ is a subgroup of $SO_{3}\left(\mathbb{R}\right)$ while $u$ defines a group homomorphism $\varphi_{u}:\:M_{g}\rightarrow SL_{3}\left(\mathbb{Q}\right)$. Let us denote $\mathrm{Im}\varphi_{u}$ as $H_{g}$. The group $M_{g}$ can be unambiguously restored from $H_{g}$ and a given matrix $u$ because $\varphi_{u}^{-1}\mid_{H_{g}}:H_{g}\rightarrow M_{g}$ is a group isomorphism. Therefore, we focus our attention on finding $H_{g}$ for any given Gram matrix. Taking into account that $r^{t}r=1$ and $ru=uh$, the equation $h^{t}gh=g$ defines the group $H_{g}$. This equation describes the indefinite orthogonal group $O\left(\mathbb{R}^{3},Q_{g}\right)$ over the real numbers with a quadratic form $Q_{g}:$ $\mathbb{R}^{3}\rightarrow\mathbb{R}$ represented by the matrix $g$. This three-parametric linear algebraic group is well studied over the field of real numbers. The specifics of our problem, however, require searching for a subgroup of all rational matrices in the group $O\left(\mathbb{R}^{3},Q_{g}\right)$ defined for an arbitrary real-valued Gram matrix $g$. 

In all cases, except for the simplest cubic lattice, this is a challenging problem. Indeed, for the cubic lattice $g=a^{2}I$, where $I$ is the identity matrix, which means that $h^{t}h=I$ and $H_{g}\cong SO_{3}\left(\mathbb{Q}\right)$ for all possible values of $a$. This result reproduces the main theorem of ref.~\cite{Wang_2024}, where 3D Moir{\'e} crystals for the cubic lattice were first introduced. However, when the Gram matrix is not proportional to the identity matrix, the parametrisation of $H_{g}$ group is much more complex.
The main difficulty comes from the fact that the elements of matrix $g$ might not belong to the field $\mathbb{Q}$. The case of a rational Gram matrix is much more straightforward as $H_{g}$ can be parametrised by the elements of Clifford algebras over $\mathbb{Q}$, as discussed in the next Section. However, the case of irrational $g$ seemingly can not be approached in the same way. The two lemmas and their proofs presented in Supplementary Text S2 allow us to build a bridge between these two cases. They show that all elements of $H_{g}$ for an irrational Gram matrix can be mapped by the elements of the corresponding groups built for rational matrices $g$.

\section*{Moir{\'e} crystals formed with rational Gram matrices}\label{sec:ration-gram}

A rational Gram matrix $g$ generates a symmetric quadratic form of the vector space $V\equiv\mathbb{Q}^{3}$ over the field of rational numbers $Q:\:V\rightarrow\mathbb{Q}$ defined as $Q\left(v\right)=\sum_{i,j}g_{ij}v_{i}v_{j}$, $\forall v\in V$. We introduce a bilinear form associated with $Q$ as $B_{Q}\left(v,u\right)=1/2\left[Q\left(v+u\right)-Q\left(v\right)-Q\left(u\right)\right]=\sum_{i,j}g_{ij}v_{i}u_{j}$, $\forall v,u\in V$. This connection between the forms $Q$ and $B_{Q}$ is unambiguous since $\mathrm{char}\left(\mathbb{Q}\right)\neq2$. Linear transformation $p$ of the vector space $V$ preserving the form $B_{Q}$ for any pair of non-zero vectors must satisfy the condition $p^{t}gp=g$, since $B_{Q}\left(v,u\right)=v^{t}gu$ and $B_{Q}\left(pv,pu\right)=\left(pv\right)^{t}gpu=v^{t}\left(p^{t}gp\right)u$. It means that a group of such linear transformations is homomorphic to the group $H_{g}$ generating all 3D Moir{\'e} lattices for $L$. It allows to reduce the task of finding the group $H_{g}$ to the problem of parametrising the group of generalised orthogonal transformations of the rational quadratic space $\left(V,Q\right)$ over the filed of rational numbers $\mathbb{Q}$. This problem can be solved with the help of the Clifford algebra $Cl\left(V,Q\right)$ associated with the form $Q$ \cite{Cassels1978}. For brevity, we will omit the vector space and quadratic form when referring to the Clifford algebra in further discussion.

Clifford algebra is an associative algebra of $\dim Cl=2^{\dim V}$ that is a quotient of the tensor algebra $T\left(V\right)$ by an ideal generated by the elements of the form $x\otimes x-Q\left(x\right)\cdot1$. $Cl$ is a $\mathbb{Z}_{2}$-graded algebra that can be decomposed into a sum of even and odd sub-algebras $Cl^{+}\oplus Cl^{-}$. Any element $p$ of $Cl$ can be expressed as $p=p_{1}\cdot\ldots\cdot p_{m}$, where $p_{i}\in V$ (more precisely, $p_{i}\in T^{1}\left(V\right)$) and $m$ is an even (odd) number for the elements of $Cl^{+}$ ($Cl^{-}$). If the elements of $Cl$ corresponding to the basis vectors of $V$ are denoted as $\left\{ \sigma_{1},\sigma_{2},\sigma_{3}\right\} $ then $Cl^{+}$ and $Cl^{-}$ have the basis sets $\left\{ 1,\sigma_{1}\sigma_{2},\sigma_{1}\sigma_{3},\sigma_{2}\sigma_{3}\right\} $ and $\left\{ \sigma_{1},\sigma_{2},\sigma_{3},\sigma_{1}\sigma_{2}\sigma_{3}\right\} $. As follows from the construction of Clifford algebra, the elements $\left\{ \sigma_{i}\right\} $ satisfy the following relationships $\sigma_{i}\sigma_{j}+\sigma_{j}\sigma_{i}=2g_{ij}$.

In general, the Gram matrix $g$ is not necessarily diagonal, however it can be transformed into diagonal form by $g=M^{t}\tilde{g}M$, where the upper-triangular matrix $M$ has the form

\begin{equation}
M=\left(\begin{array}{ccc}
1 & \frac{g_{12}}{g_{11}} & \frac{g_{13}}{g_{11}}\\
0 & 1 & \frac{g_{11}g_{23}-g_{12}g_{13}}{g_{11}g_{22}-g_{12}^{2}}\\
0 & 0 & 1
\end{array}\right)
\end{equation}

\noindent and

\begin{equation}
\tilde{g}=\mathrm{diag}\left\{ g_{11},g_{22}-\frac{g_{12}^{2}}{g_{11}},\frac{g_{11}g_{33}-g_{13}^{2}}{g_{11}}-\frac{\left(g_{11}g_{23}-g_{12}g_{13}\right)^{2}}{g_{11}\left(g_{11}g_{22}-g_{12}^{2}\right)}\right\} .
\end{equation}
\\
\noindent Here, $M$ and $\tilde{g}$ belong to $GL_{3}\left(\mathbb{Q}\right)$ and this transformation always exists because, as follows from Equation (\ref{eq:gram}), $g_{11}>0$ and $g_{11}g_{22}-g_{12}^{2}=a^{2}b^{2}\left(1-\cos^{2}\gamma\right)>0$ since the unit vectors $\mathbf{u}_{1}$ and $\mathbf{u}_{2}$ of the initial lattice $L$ are non-collinear. Group $H_{\tilde{g}}$ corresponding to the diagonal Gram matrix $\tilde{g}$ is isomorphic to the group $H_{g}$, because $H_{g}=M^{-1}H_{\tilde{g}}M$. Therefore, without losing generality, we can further consider a diagonal quadratic form $Q\left(u\right)=\widetilde{g}_{1}u_{1}^{2}+\widetilde{g}_{2}u_{2}^{2}+\widetilde{g}_{3}u_{3}^{2}$ with $\widetilde{g}_{1},\widetilde{g}_{2},\widetilde{g}_{3}\in\mathbb{Q}$.

For an invertible element $p\in Cl$ and vector $v\in T^{1}\left(V\right)$, the following linear transformation of the vector space $v\mapsto pvp^{-1}$ preserves the form $Q$ \cite{Cassels1978}. Additionally, the elements of $Cl^{+}$ correspond to linear transformations with $\det=1$, while $Cl^{-}$contains transformations having $\det=-1$. In the case of the real field, a set of elements $p\in Cl^{+}$ with a unit norm forms the Spin group of the quadratic space $\left(V,Q\right)$, which is a two-sheeted covering of $SO\left(V,Q\right)$. The same group can be constructed in the case of the rational field with the only exception that the elements can not be always normalised to unity~\cite{Cassels1978}, because $\mathbb{Q}$ is not closed with respect to the square root operation. Let us consider an element $Cl^{+}$ of the form $p=p_{0}+p_{1}\sigma_{1}\sigma_{2}+p_{2}\sigma_{1}\sigma_{3}+p_{3}\sigma_{2}\sigma_{3}$ with the inverse $p^{-1}=N^{-1}\left(p_{0}-p_{1}\sigma_{1}\sigma_{2}-p_{2}\sigma_{1}\sigma_{3}-p_{3}\sigma_{2}\sigma_{3}\right)$ with $N=p_{0}^{2}+\widetilde{g}_{1}\widetilde{g}_{2}p_{1}^{2}+\widetilde{g}_{1}\widetilde{g}_{3}p_{2}^{2}+\widetilde{g}_{2}\widetilde{g}_{3}p_{3}^{2}$. As detailed in Supplementary Materials S4, a map $\varphi_{\tilde{g}}:Cl^{+}\rightarrow H_{\tilde{g}}$ of the even component of the Clifford algebra onto group $H_{\tilde{g}}$ is given by

\begin{equation}
\varphi_{\tilde{g}}\left(p\right)=I+\frac{2}{N}\left(\begin{array}{ccc}
-\widetilde{g}_{1}\widetilde{g}_{2}p_{1}^{2}-\widetilde{g}_{1}\widetilde{g}_{3}p_{2}^{2} & \widetilde{g}_{2}\left(p_{0}p_{1}-\widetilde{g}_{3}p_{2}p_{3}\right) & \widetilde{g}_{3}\left(p_{0}p_{2}+\widetilde{g}_{2}p_{1}p_{3}\right)\\
\widetilde{g}_{1}\left(-p_{0}p_{1}-\widetilde{g}_{3}p_{2}p_{3}\right) & -\widetilde{g}_{1}\widetilde{g}_{2}p_{1}^{2}-\widetilde{g}_{2}\widetilde{g}_{3}p_{3}^{2} & \widetilde{g}_{3}\left(p_{0}p_{3}-\widetilde{g}_{1}p_{1}p_{2}\right)\\
\widetilde{g}_{1}\left(-p_{0}p_{2}+\widetilde{g}_{2}p_{1}p_{3}\right) & \widetilde{g}_{2}\left(-p_{0}p_{3}-\widetilde{g}_{1}p_{1}p_{2}\right) & -\widetilde{g}_{1}\widetilde{g}_{3}p_{2}^{2}-\widetilde{g}_{2}\widetilde{g}_{3}p_{3}^{2}
\end{array}\right).
\label{eq:param}
\end{equation}
\\

Any element different from $p$ by multiplying by a non-zero constant corresponds to the same orthogonal transformation of $\left(V,Q\right)$. Therefore, we can consider a set of the elements with $p_{0}=1$ parametrised by three coordinates $\left\{ p_{1},p_{2},p_{3}\right\} $. Equation~(\ref{eq:param}) gives a complete parametrisation of group $H_{g}$ by three rational parameters.  When $g$ matrix is equal to unity and $N=1$, Equation~(\ref{eq:param}) is transformed into the well-known parametrisation of 3D rotation matrices by quaternions. The group of rotations $M_{g}$ generating 3D Moir{\'e} lattices can be obtained from $H_{g}$ by applying isomorphism $\varphi_{u}$, as shown schematically in Fig.~\ref{fig:scheme}c. Here, we make an important side note that the proposed mathematical formalism allows us to generate a family of \textit{pseudo-Moir{\'e} lattices} corresponding to the orthogonal transformation with negative determinant (elements of $Cl^{-}$). Such 3D structures can not be obtained by a twist in real space but they still might be of interest for materials science as examples of unusual solid state phases.

\begin{figure}
	\centering
	\includegraphics[width=0.95\textwidth]{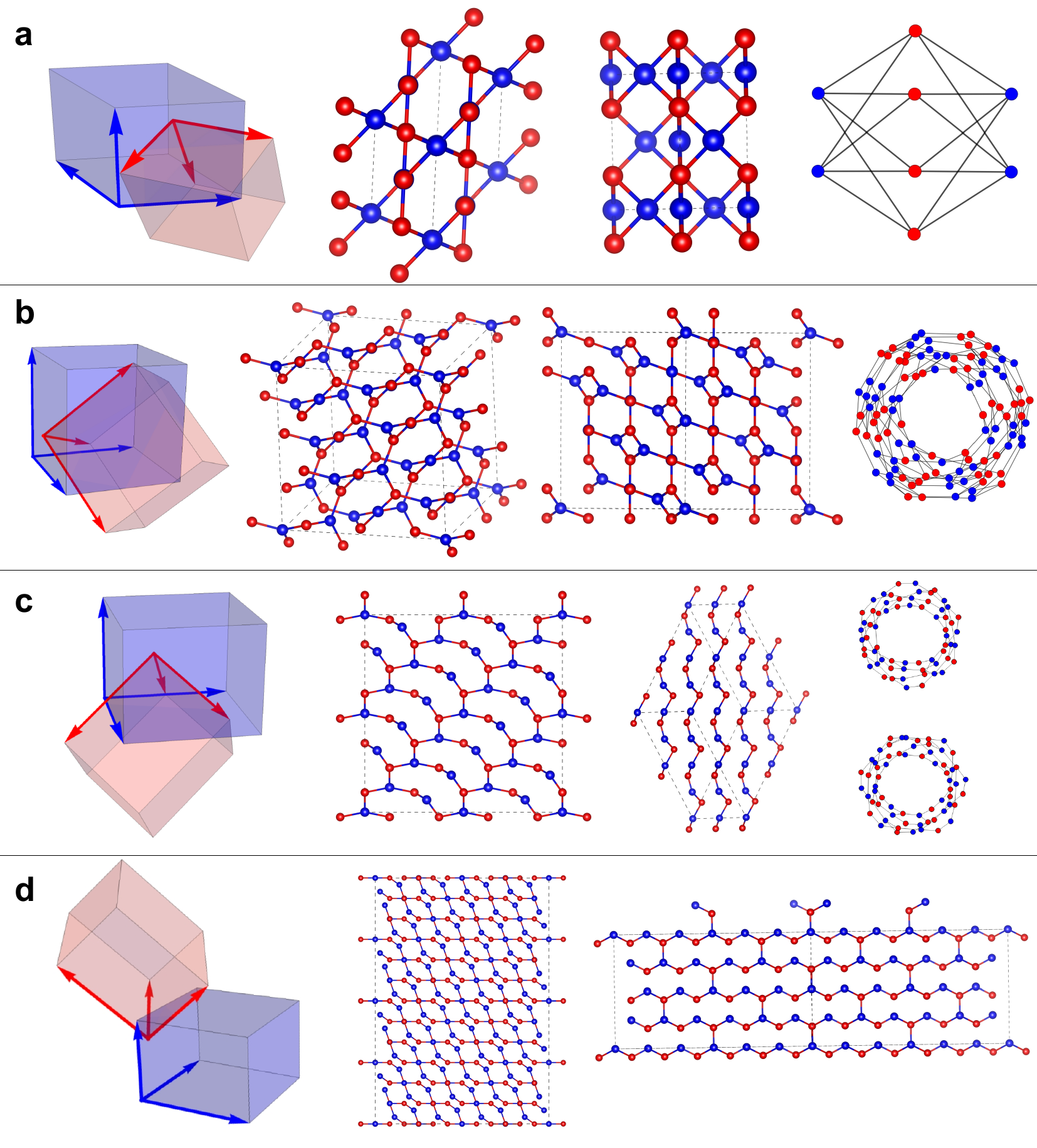}
	\caption{\textbf{Examples of three-dimensional Moir{\'e} crystals.}
		From left to right in all panels: initial unit cell together with the rotated and displaced one, two different projections of the crystal structure, fundamental graph characterising topology of the network. For the structure (\textbf{D}) the fundamental graph is not shown due to the large size of the unit cell. All crystal structures presented here are available in Supplementary Material S5.}
	\label{fig:structures}
\end{figure}

The crystal structures of some interesting examples of 3D Moir{\'e} crystals obtained with rational Gram matrices are shown in Fig.~\ref{fig:structures}, and their crystallographic properties are summarised in Table \ref{tab:summary}. General observation about all four structures is their bipartite nature due to the fact that atoms of the prototype lattice $L$ are surrounded by atoms of the lattice $rL$. This can be clearly seen from the fundamental graphs corresponding to crystals \textbf{A}-\textbf{D} shown in Fig.~\ref{fig:structures}, where every blue site is connected to red sites only and vice versa.

Structure \textbf{A} is a result of the transformation from the primitive hexagonal lattice $L$ with one atom in the unit cell to tetragonal Moir{\'e} superlattice with eight atoms in the unit cell. The unit cell vectors of \textbf{A} constructed with the algorithm described above correspond to the monoclinic system, whereas the Niggli reduced cell allows to assign this lattice to the correct tetragonal symmetry. Four atoms in the cell come from the initial lattice $L$, while the remaining four originate from $rL$. All bond lengths are equivalent in the structure \textbf{A} and every atom has a square planar coordination with a distortion of bond angles characterised by 9$^{\text{o}}$ deviation from the ideal 90$^{\text{o}}$ angle. Such coordination is typical for $d^{8}$ transition metals and is also observed in layered cuprates and mixed oxides containing iron ions~\cite{Tsujimoto_2007}.  Hence, the structure \textbf{A} can be viewed as a model of a novel 3D transition metal binary compound with square planar geometry of metal centres.

Cubic crystal \textbf{B} with 108 atoms in the unit cell is obtained from the body-centred cubic lattice $L$ containing two atoms in the unit cell. This structure gives an important example of 3D Moire superlattice belonging to the same high-symmetry crystal system as the prototype lattice $L$. It is a fully connected three-dimensional network with three-coordinated atoms arranged in 8-membered rings. Three-valence sites exhibit slightly distorted triangular coordination.

Structures \textbf{C} and \textbf{D} are layered structures, which means that their fundamental graphs are not fully connected and have several components corresponding to individual layers. In both structures two-dimensional corrugated layers are stacked on top of each other which is typical for van der Waals 2D materials. Structure \textbf{C} belongs to the tetragonal crystal system which is the same as the crystal system of the prototype lattice $L$. At the same time, in the case of \textbf{D} the resulting Moire superlattice is of lower orthorhombic symmetry as compared to the prototype tetragonal lattice. The individual layers have very unique topologies with 10- and 14-membered rings consisting of two- and three-valence sites. Their embedding in the 3D space gives interesting corrugated geometries of the layers, as demonstrated in Fig.~\ref{fig:structures}, which can be viewed as novel structural types of layered 2D materials.

\begin{table}[ht]
    \centering
	\caption{\textbf{Crystallographic properties of 3D Moir{\'e} crystals.}
		 Crystallographic properties of the prototype lattice, $L$, and the resulting 3D Moir{\'e} crystals are tabulated for systems \textbf{A}-\textbf{D} shown in Fig.~\ref{fig:structures}.
         Parameters describing the generating transformation include coordinates $(p_1,p_2,p_3)$ of the corresponding element of the Clifford algebra parametrising the rotation, Euler angles of the rotation, and the displacement vector $\mathbf{d}$ expressed in the basis of the unit vectors of Moir{\'e} crystal. Only non-trivial parameters of the unit cell  are given  assuming that $a=1$.}
	\label{tab:summary}
\setlength\tabcolsep{0pt}
\begin{tabular*}{\linewidth}{@{\extracolsep{\fill}}cccccc}
\hline 
 &  & \textbf{A} & \textbf{B} & \textbf{C} & \textbf{D}\tabularnewline
\hline 
\multirow{3}{*}{prototype lattice ($L$)} & crystal system & hexagonal & cubic & tetragonal & tetragonal\tabularnewline
 & unit cell & $c=\sqrt{3}/2$ & - & $c=\sqrt{2/3}$ & $c=1/\sqrt{2}$\tabularnewline
 & number of atoms & 1 & 2 & 2 & 4\tabularnewline
\hline 
\multirow{3}{*}{generating transformation} & $p_{1-3}$ & $\left(2,2,4/3\right)$ & $\left(1/3,1/3,-1/3\right)$ & $\left(0,3/2,3/2\right)$ & $\left(-2,1,-1\right)$\tabularnewline
 & Euler angles & $\left(60^{\textrm{o}},90^{\textrm{o}},0^{\textrm{o}}\right)$ \hfill & $\left(26.57^{\textrm{o}},48.19^{\textrm{o}},63.43^{\textrm{o}}\right)$ \hfill & $\left(135^{\textrm{o}},120^{\textrm{o}},135^{\textrm{o}}\right)$ \hfill & $\left(108.43^{\textrm{o}},48.19^{\textrm{o}},18.43^{\textrm{o}}\right)$ \tabularnewline
 & displacement & $\left(0.50,0.50,0.50\right)$ & $\left(0.00,0.33,0.33\right)$ & $\left(0.00,0.40,0.35\right)$ & $\left(0.00,0.15,0.50\right)$\tabularnewline
\hline 
\multirow{3}{*}{Moir{\'e} crystal topology} & components & 1 & 1 & 2 (layered) & 11 (layered)\tabularnewline
 & coordination & 4 & 3 & 2 and 3 & 2 and 3\tabularnewline
 & cycles & 4 and 8 & 8 & 10 & 14\tabularnewline
\hline 
\multirow{3}{*}{Moir{\'e} crystal} & crystal system & tetragonal & cubic & tetragonal & orthorombic\tabularnewline
 & unit cell & $c=1/\sqrt{3}$ & - & $c=1/\sqrt{6}$ & $b=3,\;c=3/\sqrt{2}$\tabularnewline
 & number of atoms & 8 & 108 & 128 & 576 \tabularnewline
\hline 
\end{tabular*}
\end{table}

\section*{Extension to the case of irrational Gram matrices}

From lemmas~\ref{lemma:ration-g} and \ref{lemma:centr} of Supporting Materials S2, we conclude that a rational matrix $h$ belongs to $H_{g}$ if and only if the real-valued Gram matrix $g$ can be represented as $g=g'k$, where $k\in C\left(h\right)$ belongs to the centralizer of $h$ in $GL_{3}\left(\mathbb{R}\right)$ and $g'\in GL_{3}\left(\mathbb{Q}\right)$ such that $h^{t}g'h=g'$. As $h$ is similar to the matrix of rotation $r$, it has the same characteristic polynomial and the same set of eigenvalues. Therefore, $h$ has three eigenvalues, two of which are complex: $\left\{ 1,e^{\pm i\theta}\right\} $ with $\theta\in\left[0,2\pi\right)$. If $\theta\neq0$ or $\pi$, then all three eigenvalues are distinct. The case of $\theta=0$ is trivial as it corresponds to the unity matrix, whereas the case of $\theta=\pi$ does not produce non-trivial Moir{\'e} patterns. If all three eigenvalues of the matrix $h$ are distinct, then every matrix commuting with it can be expressed as a polynomial of $h$ of degree $\dim h-1$ \cite{Gantmacher} so that

\begin{equation}
k=\kappa_{0}I+\kappa_{1}h+\kappa_{2}h^{2},
\end{equation}
\\
\noindent where $\kappa_{i}$ are, in general, complex numbers. We are interested in matrices $k$ which satisfy three additional conditions: (i) $k$ must be real; (ii) $\det k\neq0$, and (iii) the product $gk$ must be symmetric for any $g$, for which $h^{t}gh=g$. As we show in Supporting Material S2, such matrices $k$ have the following form

\begin{equation}
k=\kappa_{0}I + \kappa_{2}\left(h^{2}-2\cos\theta\cdot h\right),
\label{eq:centalizer_el}
\end{equation}

\noindent where real constants $\kappa_{0}$ and $\kappa_{2}$ satisfy the conditions $\kappa_{0}\neq\kappa_{2}$ and $\kappa_{0}\neq \kappa_{2}\left\{ 2\cos\theta-1\right\} $. We note, that the quantity $2\cos\theta$ is a coefficient of the characteristic polynomial of $h$ and, therefore, is a rational number due to rationality of $h$. It means, that the matrix $\left(h^{2}-2\cos\theta\cdot h\right)$ is rational. Hence, any irrational Gram matrix $g$ that has a non-unitary element in the group $H_{g}$ must have the form $g=\kappa_{0}g'+\kappa_{2}g''$, where both matrices $g'$ and $g''$ are rational and $\kappa_{0}$ and $\kappa_{2}$ are arbitrary real constants. This proves the following necessary condition for the existence of 3D Moir{\'e} lattice:

\begin{theorem}
\label{theorem:exist}
A non-trivial periodic 3D Moir{\'e} lattice can be constructed for a given prototype lattice $L$ only if no more than two elements of its Gram matrix are rationally independent.
\end{theorem}

This condition significantly limits a set of irrational Gram matrices that can be used to construct 3D Moir{\'e} crystals. Among six independent matrix elements determining the Gram matrix, only two can be chosen to be arbitrary irrational numbers. Other four parameters must be expressed as rational linear combinations of these two irrational numbers. For example, an orthorhombic lattice with the following Gram matrix

\begin{equation*}
g=\left(\begin{array}{ccc}
\sqrt{2} & 0 & 0\\
0 & \sqrt{3} & 0\\
0 & 0 & \pi
\end{array}\right)
\end{equation*}
\\
\noindent does not have any rotations generating non-trivial Moir{\'e} crystals, because three irrational numbers $\sqrt{2}$, $\sqrt{3}$ and $\pi$ are rationally independent -- none of them can be expressed as a linear combination of the other two with rational coefficients.

An interesting corollary to theorem \ref{theorem:exist} is the fact that any lattice $L$ belonging to either cubic, tetrahedral, rhombohedral or hexagonal crystal systems will always have non-trivial Moir{\'e} super-crystals that can be constructed from it. It comes from the specific form of their Gram matrices and the fact that the number of independent parameters defining the unit cell of these lattices does not exceed two. For other three families: orthorhombic, monoclinic and triclinic the existence of three-dimensional Moir{\'e} crystals is not guaranteed unless their Gram matrix satisfies the condition prescribed by theorem \ref{theorem:exist}.

As we established in the previous Section, the group $H_{g}$ is a three-parametric group for a rational matrix $g$. It means that this group includes elements corresponding to rotations around different non-collinear axes.  In the case of irrational Gram matrix the following statement holds:

\begin{theorem}
\label{theorem:param}
For an irrational Gram matrix $g$ satisfying the necessary condition for existence, the group $H_{g}$ is three-parametric if and only if $g=\kappa g'$, where $g'\in GL_{3}\left(\mathbb{Q}\right)$ and $\kappa>0$ is a real constant. Otherwise, the group $H_{g}$ is a one-parametric group isomorphic to a subgroup of $SO_{2}\left(\mathbb{R}\right)$.
\end{theorem}

Detailed proof of this statement is given in Supplementary Materials S3. It is obvious, that for an irrational matrix $g$ proportional to a rational matrix $g'$ the group $H_{g}$ coincides with $H_{g'}$. For such matrices the groups $H_{g}$ can be found using the methodology of the previous Section. The remaining irrational matrices of the form  $g=\kappa_{0}g'+\kappa_{2}g''$ with $\kappa_{2} \neq 0$ have one-parametric $H_{g}$ groups that are subgroups of $H_{g'}$ parametrized by Equation~(\ref{eq:param}). Geometrically, these subgroups correspond to rotations around the same axis. In conclusion, theorems \ref{theorem:exist} and \ref{theorem:param} allow us to find all irrational Gram matrices that have non-trivial periodic Moir{\'e} patterns, and reduce the problem of calculating the corresponding $H_{g}$ groups to the problem with rational Gram matrices addressed in the previous Section.

\section*{Conclusions and outlook}

This work describes the foundational principles of 3D Moir{\'e} crystallography. We present a general method of constructing 3D Moir{\'e} crystals together with their complete classification. It unlocks unprecedented opportunities for discovering novel non-trivial 3D crystal structures with diverse symmetries, topologies of chemical frameworks and tunable electronic, optical, and quantum properties which differ principally from the properties of the constituent lattices. The unmatched diversity and tunability of the potential 3D Moir{\'e} structures, most of which evade human chemical intuition, makes them a valuable tool for the community focusing on the crystal structure search and prediction.

Some transformations may generate an unphysical crystal structure but, as shown in this work, there exists a variety of realistic 3D Moir{\'e} crystals with an unexplored application potential. Developing new fabrication techniques to make 3D Moir{\'e} crystals is a major challenge, and holographic fabrication of 3D Moir{\'e} photonic crystals \cite{Hurley_2023}, using spin-dependent optical lattices in ultracold atomic gases \cite{Wang_2024}, and construction of chirality-specific material \cite{Song_2021} represent some recent advances in this newly emerging field. The general construction presented here provides a theoretical framework aiding the future development in the field of 3D Moir{\'e} materials.

\backmatter

\bmhead{Supplementary information}
Supplementary Information is available for this paper. It contains details of mathematical proofs and all crystal structures presented in the paper.

\bmhead{Acknowledgements}

The authors acknowledge financial support from two EPSRC programme grants: "Enabling Net Zero and the AI Revolution with ultra-low energy 2D Materials and Devices (NEED2D)" (UKRI-1249) and "Metal Atoms on Surfaces and Interfaces (MASI) for Sustainable Future" (EP/V000055/1).

\section*{Declarations}

\begin{itemize}
\item Funding: EPSRC programme grants: "Enabling Net Zero and the AI Revolution with ultra-low energy 2D Materials and Devices (NEED2D)" (UKRI-1249) and "Metal Atoms on Surfaces and Interfaces (MASI) for Sustainable Future" (EP/V000055/1).
\item Conflict of interest/Competing interests: there are no competing interests to declare.
\item Author contribution: IP developed the mathematical formalism of 3D Moir{\'e} crystallography and wrote the draft of the paper with further contributions from EB.
\end{itemize}



\bibliography{sn-bibliography}

\newpage


\renewcommand{\thefigure}{S\arabic{figure}}
\renewcommand{\thetable}{S\arabic{table}}
\renewcommand{\theequation}{S\arabic{equation}}
\renewcommand{\thepage}{S\arabic{page}}
\setcounter{figure}{0}
\setcounter{table}{0}
\setcounter{equation}{0}
\setcounter{page}{1} 


\begin{center}
\section*{Supplementary Information for\\ Three-dimensional Moir{\'e} crystallography}

\author{
	Ilya~Popov$^{1\ast}$,
	Elena~Besley$^{1\ast}$\\
	\small $^{1}$School of Chemistry, University of Nottingham, University Park NG7 2RD, United Kingdom.\\
	\small$^\ast$Corresponding authors. Email:
    Ilya.Popov1@nottingham.ac.uk; Elena.Besley@nottingham.ac.uk
}
\end{center}

\newtheorem{lemma}[theorem]{Lemma}
\setcounter{theorem}{0}
\renewcommand{\thetheorem}{S\arabic{theorem}}
\renewcommand{\thelemma}{S\arabic{lemma}}

\subsection*{Section S1: The group structure of $M_{g}$}


\begin{theorem}
$M_{g}$ is a subgroup of $SO_{3}\left(\mathbb{R}\right)$.
\label{theorem:mg-group}
\end{theorem}

\noindent \textit{Proof}. Consider two rotations $r_{1},r_{2}\in M_{g}$ so that matrices $h_{1}=u^{-1}r_{1}u$ and $h_{2}=u^{-1}r_{2}u$ are both rational. Matrix $h=h_{1}h_{2}=u^{-1}r_{1}r_{2}u$ is also rational, which means that $r_{1}r_{2}\in M_{g}$. Next, for any element $r\in M_{g}$ we check that $r^{-1}$ belongs to $M_{g}$: $h^{-1}=\left(u^{-1}ru\right)^{-1}=u^{-1}r^{-1}u\in SL_{3}\left(\mathbb{Q}\right)$. Finally, it is obvious, that the unity element of $SO_{3}\left(\mathbb{R}\right)$ belongs to $M_{g}$. Therefore, $M_{g}$ satisfies all axioms of the group.

\subsection*{Section S2: Mapping the elements of groups $H_{g}$ for irrational and rational Gram matrices}

\begin{lemma}
Let for a given \textit{real} non-degenerate symmetric matrix $g$ there be a matrix $h\in SL_{3}\left(\mathbb{Q}\right)$ different from the unity, such that $h^{t}gh=g$. Then there exists a \textit{rational} symmetric non-degenerate matrix $g'\in GL_{3}\left(\mathbb{Q}\right)$ satisfying the same equation.
\label{lemma:ration-g}
\end{lemma}

\noindent \textit{Proof}. The relation $h^{t}gh=g$ gives a homogeneous system of linear equations with rational coefficients with respect to the elements of the Gram matrix $g$. Indeed, simple derivations show that

\begin{equation}
\sum_{k,m}\left(h_{ki}h_{mj}-\delta_{ik}\delta_{jm}\right)g_{km}=0,\label{eq:hsle}
\end{equation}
\\
\noindent where $\delta$ is the Kronecker delta and $h_{ki},h_{mj}\in\mathbb{Q}$. If we enumerate ordered pairs of indices with a new index $\lambda=\left\{ \lambda_{1},\lambda_{2}\right\} $ and take into account that $g$ is symmetrical, we can rewrite this system as $\tilde{h}_{\lambda'\lambda}g_{\lambda}=0$ with the matrix elements
\begin{equation}
   \tilde{h}_{\lambda'\lambda}=h_{\lambda_{1}\lambda'_{1}}h_{\lambda_{2}\lambda'_{2}}-\delta_{\lambda_{1}\lambda'_{1}}\delta_{\lambda_{2}\lambda'_{2}} + (1-\delta_{\lambda_{1}\lambda_{2}})(h_{\lambda_{1}\lambda'_{2}}h_{\lambda_{2}\lambda'_{1}}-\delta_{\lambda_{1}\lambda'_{2}}\delta_{\lambda_{2}\lambda'_{1}}).
\end{equation}
\\
Note that $\tilde{h}_{\lambda'\lambda}\in\mathbb{Q}$. It is a system of six equations with six unknowns due to the symmetrical form of $g$, hence $\dim\tilde{h}=6$. If the element $h$ is different from the unity, then at least one element of the matrix $\tilde{h}$ is non-vanishing, therefore $\mathrm{rank}\left(\tilde{h}\right)>0$. At the same time, $\mathrm{rank}\left(\tilde{h}\right)<\dim\tilde{h}$, because otherwise there would be no non-zero solutions of the system $\tilde{h}_{\lambda'\lambda}g_{\lambda}=0$. Therefore, this system has $1\leq n\leq5$ independent variables, that we are going to denote as $\left\{ \kappa_{i}\right\} _{i=1\ldots n}\in\mathbb{R}$. A fundamental system of solutions of $\tilde{h}_{\lambda'\lambda}g_{\lambda}=0$ can be represented as $g_{\lambda}=\sum_{i=1,n}c_{\lambda i}\kappa_{i}$ with $c_{\lambda i}\in\mathbb{Q}$. It means, that the Gram matrix satisfying $h^{t}gh=g$ has the form $g=\sum_{i=1,n}\kappa_{i}g_{i}$, where $g_{i}$ are rational matrices, that individually satisfy $h^{t}g_{i}h=g_{i}$. For any set of rational coefficients $\kappa'_{i}$ the rational symmetric matrix $g'=\sum_{i=1,n}\kappa'_{i}g_{i}$ will satisfy the equation $h^{t}g'h=g'$. 

We now need to prove that there exist a set of \textit{rational} coefficients $\kappa'_{i}$ such that $\det g' \neq 0$. By the condition of the theorem $\det g \neq 0$, which means that there is a set of \textit{real} coefficients $\kappa_{i}$ such that corresponding matrix $g$ is non-degenerate. Based on the Hurwitz theorem, for every irrational $\kappa_{i}$ there exist infinitely many pairs of integers $m_{i}$, $l_{i}$ such that

\begin{equation}
    \left|\kappa_{i}-\frac{m_{i}}{l_{i}}\right|<\frac{1}{\sqrt{5}l_{i}^{2}}.
    \label{eq:hurwitz}
\end{equation}
\\
At the same time, determinant of matrix $g$ can be expressed through the coefficients $\kappa_{i}$ as

\begin{equation}
    \det g=\sum c_{i_{1}i_{2}i_{3}}\cdot\kappa_{i_{1}}\kappa_{i_{2}}\kappa_{i_{3}},
    \label{eq:det}
\end{equation}
\\
\noindent where rational coefficients $c_{i_{1}i_{2}i_{3}}$ have the form

\begin{equation}
    c_{i_{1}i_{2}i_{3}}=\sum_{\alpha_{1},\alpha_{2},\alpha_{3}}\varepsilon_{\alpha_{1}\alpha_{2}\alpha_{3}}\left(g_{i_{1}}\right)_{1\alpha_{1}}\left(g_{i_{2}}\right)_{1\alpha_{2}}\left(g_{i_{3}}\right)_{1\alpha_{3}}.
\end{equation}
\\
Setting $\kappa'_{i}=m_{i}/l_{i}$ and combining Equations~(\ref{eq:hurwitz}) and (\ref{eq:det}) we get the following inequality

\begin{equation}
    \det g-\frac{c_{2}}{\sqrt{5}l}<\det g'<\det g+\frac{c_{2}}{\sqrt{5}l},
\end{equation}
\\
where $l=\min\left\{m_{i}l_{i}\right\}$, $c_{1}$ and $c_{2}$ are positive finite constants. From this inequality and the Hurwitz theorem we conclude that there always exists a set of rational coefficients $\kappa'_{i}$ so that $\det g' \neq 0$.

\begin{lemma}
Consider non-degenerate Gram matrices $g_{1}$ and $g_{2}$ and their corresponding groups $H_{g_{1}}$ and $H_{g_{2}}$. Then $g_{1}^{-1}g_{2}$ and $g_{2}^{-1}g_{1}\in C\left(H_{g_{1}}\cap H_{g_{2}}\right)$, where $C$ is the centralizer of the group $H_{g_{1}}\cap H_{g_{2}}$ in $GL_{3}\left(\mathbb{R}\right)$.
\label{lemma:centr}
\end{lemma}

\noindent \textit{Proof}. Consider an element $h\in H_{g_{1}}\cap H_{g_{2}}$. From $h^{t}g_{1}h=g_{1}$ and $\det g_{1}\neq0$ we derive that $h^{-1}g_{1}^{-1}h^{-t}=g_{1}^{-1}$. At the same time, $h^{t}g_{2}h=g_{2}$. Multiplying the former equation by the latter we get $h^{-1}g_{1}^{-1}g_{2}h=g_{1}^{-1}g_{2}$, which is equivalent to $g_{1}^{-1}g_{2}h=hg_{1}^{-1}g_{2}$. Therefore, $g_{1}^{-1}g_{2}\in C\left(H_{g_{1}}\cap H_{g_{2}}\right)$ Taking into account that $C$ is a group, $\left(g_{1}^{-1}g_{2}\right)^{-1}=g_{2}^{-1}g_{1}$ also belongs to it.

\noindent \textbf{Details required for the proof of theorem~\ref{theorem:exist} of the main text.} As discussed in the main text, we look for the elements $k\in C\left(h\right)$ for which $gk$ is symmetric and $\det k \neq 0$. It means that $gk=\left(gk\right)^{t}=k^{t}g^{t}$, which rewrites as

\begin{equation}
\kappa_{1}gh+\kappa_{2}gh^{2}=\kappa_{1}h^{t}g+\kappa_{2}\left(h^{t}\right)^{2}g.
\end{equation}
\\
Multiplying both sides on the right by $h^{2}$ and taking into account that $h^{t}gh=g$ we obtain the following

\begin{equation}
\kappa_{1}gh^{3}+\kappa_{2}gh^{4}=\kappa_{1}gh+\kappa_{2}g.
\end{equation}
\\
This is equivalent to the following matrix polynomial

\begin{equation}
\left(h+I\right)\left(h-I\right)\left(\kappa_{2}h^{2}+\kappa_{1}h+\kappa_{2}\right)=0.
\end{equation}
\\
The polynomial on the left is an annihilating polynomial of the matrix $h$, therefore, it has to be divisible by its minimal polynomial, which is the same as characteristic polynomial in our case. The characteristic polynomial of $h$ has the following form $(\lambda-1)(\lambda^{2}-2\cos\theta \cdot \lambda + 1)$ It means that $\kappa_{1}=-2\kappa_{2}\cos\theta$. This allows us to obtain Equation~(\ref{eq:centalizer_el}) of the main text.

Matrix $h$ can be expressed as $h=ph_{d}p^{-1}$, where $h_{d}=\mathrm{diag}\left\{ 1,e^{i\theta},e^{-i\theta}\right\}$ and columns of $p$ are the right eigenvectors of $h$, whereas rows of $p^{-1}$ are the left eigenvectors of $h$. This allows to express $k$ as

\begin{equation}
k=\kappa_{2}p(h_{d}^{2}-2\cos\theta\cdot h_{d}+\frac{\kappa_{0}}{\kappa_{2}}I)p^{-1},
\label{eq:k_hd}
\end{equation}
\\
where

\begin{equation}
h_{d}^{2}-2\cos\theta\cdot h_{d}+\frac{\kappa_{0}}{\kappa_{2}}I=\mathrm{diag}\left\{ 1-2\cos\theta+\frac{\kappa_{0}}{\kappa_{2}},\frac{\kappa_{0}}{\kappa_{2}}-1,\frac{\kappa_{0}}{\kappa_{2}}-1\right\} .\label{eq:diag_centr}
\end{equation}
\\
This can be rewritten as

\begin{equation}
h_{d}^{2}-2\cos\theta\cdot h_{d}+\frac{\kappa_{0}}{\kappa_{2}}I=\left(\frac{\kappa_{0}}{\kappa_{2}}-1\right)I+\left(2-2\cos\theta\right)\cdot\mathrm{diag}\left\{ 1,0,0\right\} .
\end{equation}
\\
Substituting this expression in Equation (\ref{eq:k_hd}) gives the following

\begin{equation}
k=\left(\kappa_{0}-\kappa_{2}\right)I+\kappa_{2}\left(2-2\cos\theta\right)\cdot p \cdot \mathrm{diag}\left\{ 1,0,0\right\} \cdot p^{-1}.
\end{equation}
\\
If we denote right and left eigenvectors of $h$ corresponding to the eigenvalue $\lambda=1$ as $w_{r}$ and $w_{l}$ , we get

\begin{equation}
k=\kappa'_{0}I+\kappa'_{2}\cdot w_{r}\otimes w_{l},\label{eq:k-eigenv}
\end{equation}
\\
where $\kappa'_{0}=\kappa_{0}-\kappa_{2}$, $\kappa'_{2}=\kappa_{2}\left(2-2\cos\theta\right)$ and $\otimes$ is the outer product of vectors. We note, that Equation (\ref{eq:diag_centr}) allows to express determinant of $k$ as 

\begin{equation}
   \det k=\left(\kappa_{0}-\kappa_{2}\right)^{2}\left(\kappa_{0}+\kappa_{2}\left\{ 1-2\cos\theta\right\} \right). 
\end{equation}
\\
As it is mentioned in the main text, we are only interested in $k\left(h\right)$ matrices with non-zero determinant. It means that $\kappa_{0}\neq\kappa_{2}$ and $\kappa_{0}\neq \kappa_{2}\left\{ 2\cos\theta-1\right\} $.

\subsection*{Section S3: Proof of theorem \ref{theorem:param} of the main text}

Let us consider two elements $h_{1}$ and $h_{2}$ belonging to $H_{g}\cap H_{g'}$, where $g'$ is a non-degenerate rational matrix. Based on lemma \ref{lemma:centr}, $g=g'k\left(h_{1}\right)$ and $g=g'k\left(h_{2}\right)$, which means that $k\left(h_{1}\right)=k\left(h_{2}\right)$. Using Equation (\ref{eq:k-eigenv}) this expression can be rewritten as

\begin{equation}
\kappa{}_{0}I+\kappa{}_{2}\cdot w_{r}\otimes w_{l}=\kappa'_{0}I+\kappa'_{2}\cdot v_{r}\otimes v_{l},
\end{equation}
\\
where $w_{r}$ ($w_{l}$) and $v_{r}$ ($v_{l}$) are the right (left) eigenvectors of matrices $h_{1}$ and $h_{2}$ respectively, all corresponding to the eigenvalue $\lambda=1$. Let us assume that $\kappa{}_{0}\neq\kappa'_{0}$. In this case, $\mathrm{rank}\left(\kappa_{0}I-\kappa'_{0}I\right)=3$ which means that $\mathrm{rank}\left(\kappa'_{2}\cdot v_{r}\otimes v_{l}-\kappa_{2}\cdot w_{r}\otimes w_{l}\right)=3$. The latter, however, is not possible, because $\mathrm{rank}\left(w_{r}\otimes w_{l}\right)=\mathrm{rank}\left(v_{r}\otimes v_{l}\right)=1$ and

\begin{equation}
\mathrm{rank}\left(\kappa'_{2}\cdot v_{r}\otimes v_{l}-\kappa{}_{2}\cdot w_{r}\otimes w_{l}\right)\leq\mathrm{rank}\left(\kappa{}_{2}\cdot w_{r}\otimes w_{l}\right)+\mathrm{rank}\left(\kappa'_{2}\cdot v_{r}\otimes v_{l}\right)=2.
\end{equation}
\\
Therefore, $\kappa_{0}=\kappa'_{0}$ and $\kappa_{2}\cdot w_{r}\otimes w_{l}=-\kappa'_{2}\cdot v_{r}\otimes v_{l}$. We remind that every matrix $h$ is connected to a matrix of rotation $r$ as $h=u^{-1}ru$. It means that we have the following relations for the right and left eigenvectors: $r\left(uw_{r}\right)=uw_{r}$ and $\left(w_{l}^{t}u^{-1}\right)r=w_{l}^{t}u^{-1}$. Since matrices $r$ and $r^{t}=r^{-1}$ correspond to rotation around the same axis, their eigenvectors are proportional to each other. Hence, the right and left eigenvectors of $r$ corresponding to eigenvalue $\lambda=1$ are proportional to one another. Let us denote this eigenvector as $z$. We can express eigenvectors of $h$ through the eigenvector of $r$ as $w_{r}=u^{-1}z$ and $w_{l}^{t}=z^{t}u$. Moreover, $w_{r}\otimes w_{l}=w_{r}w_{l}^{t}=u^{-1}zz^{t}u$.

If matrices $h_{1}$ and $h_{2}$ correspond to rotations $r_{1}$ and $r_{2}$, then from the equation $\kappa_{2}\cdot w_{r}\otimes w_{l}=-\kappa'_{2}\cdot v_{r}\otimes v_{l}$ we obtain that $z_{1}\otimes z_{1}\propto z_{2}\otimes z_{2}$. This means that rotations $r_{1}$ and $r_{2}$ have the same axis and belong to the same subgroup $SO_{2}\left(\mathbb{R}\right)$. Therefore, if the Gram matrix $g$ is not proportional to a rational matrix, then all elements of $H_{g}\cap H_{g'}$ correspond to the rotation around the same axis.

Finally, let us consider if it is possible that there exist two rational non-degenerate matrices $g_{1}$ and $g_{2}$ such that the groups $H_{g}\cap H_{g_{1}}$ and $H_{g}\cap H_{g_{2}}$ are both non-trivial and do not coincide. Let us consider two elements $h_{1}\in H_{g}\cap H_{g_{1}}$ such that $h_{1}\notin H_{g}\cap H_{g_{2}}$ and $h_{2}\in H_{g}\cap H_{g_{2}}$. Based on lemmas \ref{lemma:ration-g} and \ref{lemma:centr}, we write

\begin{equation}
g=g_{1}\left(\kappa_{0}I+\kappa_{2}\left\{ h_{1}^{2}-2\cos\theta_{1}\cdot h_{1}\right\} \right),\label{eq:g_g1}
\end{equation}
\\
\begin{equation}
g=g_{2}\left(\kappa'_{0}I+\kappa'_{2}\left\{ h_{2}^{2}-2\cos\theta_{2}\cdot h_{2}\right\} \right),\label{eq:g_g2}
\end{equation}
\\
where $\kappa_{0,2}$ and $\kappa'_{0,2}$ are irrational numbers. Elements of matrix $g$ are linear combination of two irrational numbers $\kappa_{0}$ and $\kappa_{2}$ with rational coefficients. Therefore, coefficients $\kappa'_{0}$ and $\kappa'_{2}$ must be rationally dependant on these numbers

\begin{equation}
\begin{array}{cc}
\kappa'_{0}=q_{00}\kappa{}_{0}+q_{02}\kappa{}_{2},\\
\kappa'_{2}=q_{20}\kappa{}_{0}+q_{22}\kappa{}_{2},
\end{array}
\end{equation}
\\
where $q_{ij}\in\mathbb{Q}$. Substituting these expressions in Equation (\ref{eq:g_g2}) we obtain

\begin{equation}
g=\kappa{}_{0}g_{2}\left(q_{00}+q_{20}\left\{ h_{2}^{2}-2\cos\theta_{2}\cdot h_{2}\right\} \right)+\kappa{}_{2}g_{2}\left(q_{02}+q_{22}\left\{ h_{2}^{2}-2\cos\theta_{2}\cdot h_{2}\right\} \right).\label{eq:g_g2-1}
\end{equation}
\\
Rational matrices in front of the coefficients $\kappa_{0}$ and $\kappa_{2}$ in Equation (\ref{eq:g_g1}) and the latter expression we get the following system of equations

\begin{equation}
\begin{array}{cc}
     g_{2}\left(q_{00}+q_{20}\left\{ h_{2}^{2}-2\cos\theta_{2}\cdot h_{2}\right\} \right)=g_{1}  \\
     g_{2}\left(q_{02}+q_{22}\left\{ h_{2}^{2}-2\cos\theta_{2}\cdot h_{2}\right\} \right)=g_{1}\left\{ h_{1}^{2}-2\cos\theta_{1}\cdot h_{1}\right\}. 
\end{array}.
\end{equation}
\\
This allows to express $g_{2}$ as

\begin{equation}
g_{2}=g_{1}\left(q_{00}q_{22}-q_{02}q_{20}\right)^{-1}\left(q_{22}I-q_{20}\left\{ h_{1}^{2}-2\cos\theta_{1}\cdot h_{1}\right\} \right).
\end{equation}
\\
It means that $h_{1}\in H_{g}\cap H_{g_{2}}$, and the groups $H_{g}\cap H_{g_{1}}$ and $H_{g}\cap H_{g_{2}}$ coincide if both are non-trivial.

\subsection*{Section S4: Connection between elements of $Cl^{+}\left(\mathbb{Q}^{3},Q\right)$
and the group $H_{g}$} \label{si-cliff}

As discussed in the main text, it is sufficient to consider the case of diagonal quadratic forms $Q\left(v\right)=g_{1}v_{1}^{2}+g_{2}v_{2}^{2}+g_{3}v_{3}^{2}$. Note, that unlike the main text, this Section employs notations of the diagonal elements of $g$ without tildes to avoid cumbersome equations. Let us consider an element of $Cl^{+}\left(\mathbb{Q}^{3},Q\right)$ having the form $p=p_{0}+p_{1}\sigma_{1}\sigma_{2}+p_{2}\sigma_{1}\sigma_{3}+p_{3}\sigma_{2}\sigma_{3}$. Its inverse is expressed as $p^{-1}=N^{-1}\left(p_{0}-p_{1}\sigma_{1}\sigma_{2}-p_{2}\sigma_{1}\sigma_{3}-p_{3}\sigma_{2}\sigma_{3}\right)$ with $N=p_{0}^{2}+g_{1}g_{2}p_{1}^{2}+g_{1}g_{3}p_{2}^{2}+g_{2}g_{3}p_{3}^{2}$. Denoting the non-scalar part as $q=p_{1}\sigma_{1}\sigma_{2}+p_{2}\sigma_{1}\sigma_{3}+p_{3}\sigma_{2}\sigma_{3}$, we rewrite $p=p_{0}+q$ and $p^{-1}=N^{-1}\left(p_{0}-q\right)$. We also remind that the generators $\sigma_{1-3}$ satisfy the following anti-commutation relations: $\sigma_{i}\sigma_{j}+\sigma_{j}\sigma_{i}=2g_{i}\delta_{ij}$. 

\noindent First of all, it is useful to check that our expression for the inverse of $p$ is correct

\begin{equation}
    pp^{-1}=N^{-1}\left(p_{0}^{2}-q^{2}\right),
\end{equation}

\noindent where

\begin{multline}
q^{2}=-g_{1}g_{2}p_{1}^{2}-g_{1}g_{3}p_{2}^{2}-g_{2}g_{3}p_{3}^{2}+p_{1}p_{2}\left(\sigma_{1}\sigma_{2}\sigma_{1}\sigma_{3}+\sigma_{1}\sigma_{3}\sigma_{1}\sigma_{2}\right)+\\
+p_{1}p_{3}\left(\sigma_{1}\sigma_{2}\sigma_{2}\sigma_{3}+\sigma_{2}\sigma_{3}\sigma_{1}\sigma_{2}\right)+p_{2}p_{3}\left(\sigma_{1}\sigma_{3}\sigma_{2}\sigma_{3}+\sigma_{2}\sigma_{3}\sigma_{1}\sigma_{3}\right).
\end{multline}
\\
All expressions in the brackets entering the last equation are equal to zero due to the anti-commutation relations. Indeed,

\[
\sigma_{1}\sigma_{2}\sigma_{1}\sigma_{3}+\sigma_{1}\sigma_{3}\sigma_{1}\sigma_{2}=-g_{1}\sigma_{2}\sigma_{3}-g_{1}\sigma_{3}\sigma_{2}=0,
\]

\[
\sigma_{1}\sigma_{2}\sigma_{2}\sigma_{3}+\sigma_{2}\sigma_{3}\sigma_{1}\sigma_{2}=g_{2}\sigma_{1}\sigma_{3}+g_{2}\sigma_{3}\sigma_{1}=0,
\]

\[
\sigma_{1}\sigma_{3}\sigma_{2}\sigma_{3}+\sigma_{2}\sigma_{3}\sigma_{1}\sigma_{3}=-g_{3}\sigma_{1}\sigma_{2}-g_{3}\sigma_{2}\sigma_{1}=0.
\]
\\
Therefore, we get $q^{2}=-g_{1}g_{2}p_{1}^{2}-g_{1}g_{3}p_{2}^{2}-g_{2}g_{3}p_{3}^{2}$ and $pp^{-1}=N^{-1}\left(p_{0}^{2}-q^{2}\right)=1$.

\noindent Let us now consider a vector $v=v_{1}\sigma_{1}+v_{2}\sigma_{2}+v_{3}\sigma_{3}$ and a linear transformation $\psi_{p}$: $v\mapsto pvp^{-1}$. Transformation $\psi_{p}$ acts on the components $\sigma_{i}$ as

\begin{equation}
\psi_{p}\left(\sigma_{i}\right)=\sum_{j=1}^{3}c_{ij}\sigma_{j}.\label{eq:expand-basis}
\end{equation}
\\
It means that the vector $v$ is transformed as

\begin{equation}
\psi_{p}\left(v\right)=\sum_{j=1}^{3}\sigma_{j}\sum_{i=1}^{3}v_{i}c_{ij}=\sum_{j=1}^{3}\bar{v}_{j}\sigma_{j},
\end{equation}

\noindent where $\bar{v}_{j}=\sum_{i=1}^{3}v_{i}c_{ij}$ are coordinates of the vector $\psi_{p}\left(v\right)$ in the basis set $\left\{ \sigma_{j}\right\} $. In matrix notations we rewrite

\begin{equation}
\left(\begin{array}{c}
\bar{v}_{1}\\
\bar{v}_{2}\\
\bar{v}_{3}
\end{array}\right)=\left(\begin{array}{ccc}
c_{11} & c_{21} & c_{31}\\
c_{12} & c_{22} & c_{32}\\
c_{13} & c_{23} & c_{33}
\end{array}\right)\left(\begin{array}{c}
v_{1}\\
v_{2}\\
v_{3}
\end{array}\right).\label{eq:v-transform}
\end{equation}

The transformation matrix in the last equation is the element of $H_{g}$ corresponding to the element $p$ of the Clifford algebra. Therefore, our goal is to determine coefficients $c_{ij}$ . To do that let is consider vectors $\psi_{p}\left(\sigma_{i}\right)=p\sigma_{i}p^{-1}$ in detail. By substituting expressions for $p$ and $p^{-1}$ we find

\begin{equation}
p\sigma_{i}p^{-1}=N^{-1}\left(p_{0}^{2}\sigma_{i}+p_{0}\left[q,\sigma_{i}\right]-q\sigma_{i}q\right),
\end{equation}

\noindent where $\left[\right]$ stands for the commutator. Taking into account that $q\sigma_{i}=\left[q,\sigma_{i}\right]+\sigma_{i}q$ we obtain

\begin{equation}
p\sigma_{i}p^{-1}=N^{-1}\left(\sigma_{i}\left\{ p_{0}^{2}-q^{2}\right\} +\left[q,\sigma_{i}\right]\left\{ p_{0}-q\right\} \right)=\sigma_{i}+N^{-1}\left[q,\sigma_{i}\right]\left\{ p_{0}-q\right\} .\label{eq:basis-transf}
\end{equation}
\\
Using the anti-commutation relations between the generators, we evaluate three commutators:

\begin{multline*}
\left[q,\sigma_{1}\right]=p_{1}\sigma_{1}\sigma_{2}\sigma_{1}+p_{2}\sigma_{1}\sigma_{3}\sigma_{1}+p_{3}\sigma_{2}\sigma_{3}\sigma_{1}-p_{1}\sigma_{1}\sigma_{1}\sigma_{2}-p_{2}\sigma_{1}\sigma_{1}\sigma_{3}-p_{3}\sigma_{1}\sigma_{2}\sigma_{3}=\\
=-2g_{1}\left(p_{1}\sigma_{2}+p_{2}\sigma_{3}\right),
\end{multline*}

\begin{multline*}
\left[q,\sigma_{2}\right]=p_{1}\sigma_{1}\sigma_{2}\sigma_{2}+p_{2}\sigma_{1}\sigma_{3}\sigma_{2}+p_{3}\sigma_{2}\sigma_{3}\sigma_{2}-p_{1}\sigma_{2}\sigma_{1}\sigma_{2}-p_{2}\sigma_{2}\sigma_{1}\sigma_{3}-p_{3}\sigma_{2}\sigma_{2}\sigma_{3}=\\
=2g_{2}\left(p_{1}\sigma_{1}-p_{3}\sigma_{3}\right),
\end{multline*}

\begin{multline*}
\left[q,\sigma_{3}\right]=p_{1}\sigma_{1}\sigma_{2}\sigma_{3}+p_{2}\sigma_{1}\sigma_{3}\sigma_{3}+p_{3}\sigma_{2}\sigma_{3}\sigma_{3}-p_{1}\sigma_{3}\sigma_{1}\sigma_{2}-p_{2}\sigma_{3}\sigma_{1}\sigma_{3}-p_{3}\sigma_{3}\sigma_{2}\sigma_{3}=\\
=2g_{3}\left(p_{2}\sigma_{1}+p_{3}\sigma_{2}\right).
\end{multline*}
\\
We also need to evaluate elements $\left[q,\sigma_{i}\right]q$. For $\left[q,\sigma_{1}\right]q$ we get

\begin{multline*}
\left[q,\sigma_{1}\right]q=-2g_{1}\left(p_{1}\sigma_{2}q+p_{2}\sigma_{3}q\right)=\\
=-2g_{1}\left(p_{1}\sigma_{2}\left\{ p_{1}\sigma_{1}\sigma_{2}+p_{2}\sigma_{1}\sigma_{3}+p_{3}\sigma_{2}\sigma_{3}\right\} +p_{2}\sigma_{3}\left\{ p_{1}\sigma_{1}\sigma_{2}+p_{2}\sigma_{1}\sigma_{3}+p_{3}\sigma_{2}\sigma_{3}\right\} \right)=\\
=-2g_{1}\left(-g_{2}p_{1}^{2}\sigma_{1}+g_{2}p_{1}p_{3}\sigma_{3}-g_{3}p_{2}^{2}\sigma_{1}-g_{3}p_{2}p_{3}\sigma_{2}\right).
\end{multline*}
\\
For $\left[q,\sigma_{2}\right]q$ we get the following

\begin{multline*}
\left[q,\sigma_{2}\right]q=2g_{2}\left(p_{1}\sigma_{1}q-p_{3}\sigma_{3}q\right)=\\
=2g_{2}\left(p_{1}\sigma_{1}\left\{ p_{1}\sigma_{1}\sigma_{2}+p_{2}\sigma_{1}\sigma_{3}+p_{3}\sigma_{2}\sigma_{3}\right\} -p_{3}\sigma_{3}\left\{ p_{1}\sigma_{1}\sigma_{2}+p_{2}\sigma_{1}\sigma_{3}+p_{3}\sigma_{2}\sigma_{3}\right\} \right)=\\
=2g_{2}\left(g_{1}p_{1}^{2}\sigma_{2}+g_{1}p_{1}p_{2}\sigma_{3}+g_{3}p_{2}p_{3}\sigma_{1}+g_{3}p_{3}^{2}\sigma_{2}\right).
\end{multline*}
\\
For $\left[q,\sigma_{3}\right]q$ we get the following

\begin{multline*}
\left[q,\sigma_{3}\right]q=2g_{3}\left(p_{2}\sigma_{1}q+p_{3}\sigma_{2}q\right)=\\
=2g_{3}\left(p_{2}\sigma_{1}\left\{ p_{1}\sigma_{1}\sigma_{2}+p_{2}\sigma_{1}\sigma_{3}+p_{3}\sigma_{2}\sigma_{3}\right\} +p_{3}\sigma_{2}\left\{ p_{1}\sigma_{1}\sigma_{2}+p_{2}\sigma_{1}\sigma_{3}+p_{3}\sigma_{2}\sigma_{3}\right\} \right)=\\
=2g_{3}\left(g_{1}p_{1}p_{2}\sigma_{2}+g_{1}p_{2}^{2}\sigma_{3}-g_{2}p_{1}p_{3}\sigma_{1}+g_{2}p_{3}^{2}\sigma_{3}\right).
\end{multline*}
\\
Substituting expressions for $\left[q,\sigma_{i}\right]q$ into equation (\ref{eq:basis-transf}) we obtain the following expansions of the vectors $\psi_{p}\left(\sigma_{i}\right)=p\sigma_{i}p^{-1}$ in the basis of $\left\{ \sigma_{i}\right\} $:

\begin{multline*}
\psi_{p}\left(\sigma_{1}\right)=\sigma_{1}+2g_{1}N^{-1}\left\{ \left(-g_{2}p_{1}^{2}-g_{3}p_{2}^{2}\right)\sigma_{1}+\left(-p_{0}p_{1}-g_{3}p_{2}p_{3}\right)\sigma_{2}+\left(-p_{0}p_{2}+g_{2}p_{1}p_{3}\right)\sigma_{3}\right\} ,\\
\psi_{p}\left(\sigma_{2}\right)=\sigma_{2}+2g_{2}N^{-1}\left\{ \left(p_{0}p_{1}-g_{3}p_{2}p_{3}\right)\sigma_{1}+\left(-g_{1}p_{1}^{2}-g_{3}p_{3}^{2}\right)\sigma_{2}+\left(-p_{0}p_{3}-g_{1}p_{1}p_{2}\right)\sigma_{3}\right\} ,\\
\psi_{p}\left(\sigma_{3}\right)=\sigma_{3}+2g_{3}N^{-1}\left\{ \left(p_{0}p_{2}+g_{2}p_{1}p_{3}\right)\sigma_{1}+\left(p_{0}p_{3}-g_{1}p_{1}p_{2}\right)\sigma_{2}+\left(-g_{1}p_{2}^{2}-g_{2}p_{3}^{2}\right)\sigma_{3}\right\} .
\end{multline*}
\\
From these expansions and equations (\ref{eq:expand-basis}, \ref{eq:v-transform}) one can easily derive the Equation (\ref{eq:param}) of the main text.

\subsection*{Section S5: Crystal structures A-D}

Here we provide crystal structures of 3D Moir{\'e} crystals \textbf{A}-\textbf{D} described in the main text. We provide the unit cell vectors and fractional coordinates of all atoms in the unit cell. Atoms originating from the prototype lattice $L$ are denoted by the symbol ``O'', while atoms originating from the lattice $rL$ are denoted by ``B''. All structures are scaled so that minimum distance between atoms always equals to 1.5 {\AA}. This is done for visualisation purposes. The presented structures can be visualised by VESTA~\cite{Momma_2011} or any other software supporting visualisation of crystallographic data.

\noindent\makebox[\linewidth]{\rule{16.5cm}{0.4pt}}

\begin{verbatim}
Structure A; hP(3/4;2/1,2/1,4/3) shift = 0.50000   0.50000   0.50000
1.0
-2.2677991390         0.0000000000         0.0000000000
0.0000183063         0.0000005230         3.9279327392
-0.0000000000         3.9279258251         0.0000000000
O    B
4    4
Direct
     0.750000000         0.250000000         0.250000000
     0.750000000         0.750000000         0.250000000
     0.250000000         0.250000000         0.750000000
     0.250000000         0.750000000         0.750000000
     0.500000000         0.500000000         0.000000000
     0.000000000         0.000000000         0.500000000
     0.000000000         0.000000000         0.000000000
     0.500000000         0.500000000         0.500000000
\end{verbatim}

\noindent\makebox[\linewidth]{\rule{16.5cm}{0.4pt}}

\begin{verbatim}
StructureB; cI(1/3,1/3,-1/3) shift = 0.00000   0.33000   0.33000
3.21412
2.00000  -1.00000  -2.00000
2.00000   2.00000   1.00000
1.00000  -2.00000   2.00000
O   B
54  54
Direct
   0.00000   0.11000   0.11000
   0.16667   0.27667   0.27667
   0.00000   0.11000   0.44333
   0.16667   0.27667   0.61000
   0.00000   0.11000   0.77667
   0.16667   0.27667   0.94333
   0.00000   0.44333   0.11000
   0.16667   0.61000   0.27667
   0.00000   0.44333   0.44333
   0.16667   0.61000   0.61000
   0.00000   0.44333   0.77667
   0.16667   0.61000   0.94333
   0.00000   0.77667   0.11000
   0.16667   0.94333   0.27667
   0.00000   0.77667   0.44333
   0.16667   0.94333   0.61000
   0.00000   0.77667   0.77667
   0.16667   0.94333   0.94333
   0.33333   0.11000   0.11000
   0.50000   0.27667   0.27667
   0.33333   0.11000   0.44333
   0.50000   0.27667   0.61000
   0.33333   0.11000   0.77667
   0.50000   0.27667   0.94333
   0.33333   0.44333   0.11000
   0.50000   0.61000   0.27667
   0.33333   0.44333   0.44333
   0.50000   0.61000   0.61000
   0.33333   0.44333   0.77667
   0.50000   0.61000   0.94333
   0.33333   0.77667   0.11000
   0.50000   0.94333   0.27667
   0.33333   0.77667   0.44333
   0.50000   0.94333   0.61000
   0.33333   0.77667   0.77667
   0.50000   0.94333   0.94333
   0.66667   0.11000   0.11000
   0.83333   0.27667   0.27667
   0.66667   0.11000   0.44333
   0.83333   0.27667   0.61000
   0.66667   0.11000   0.77667
   0.83333   0.27667   0.94333
   0.66667   0.44333   0.11000
   0.83333   0.61000   0.27667
   0.66667   0.44333   0.44333
   0.83333   0.61000   0.61000
   0.66667   0.44333   0.77667
   0.83333   0.61000   0.94333
   0.66667   0.77667   0.11000
   0.83333   0.94333   0.27667
   0.66667   0.77667   0.44333
   0.83333   0.94333   0.61000
   0.66667   0.77667   0.77667
   0.83333   0.94333   0.94333
   0.05556   0.05556   0.27778
   0.00000   0.00000   0.00000
   0.38889   0.05556   0.61111
   0.16667   0.16667   0.83333
   0.50000   0.16667   0.16667
   0.33333   0.00000   0.33333
   0.27778   0.27778   0.38889
   0.11111   0.11111   0.55556
   0.05556   0.38889   0.61111
   0.22222   0.22222   0.11111
   0.16667   0.50000   0.16667
   0.00000   0.33333   0.33333
   0.72222   0.05556   0.94444
   0.83333   0.16667   0.50000
   0.66667   0.00000   0.66667
   0.61111   0.27778   0.72222
   0.44444   0.11111   0.88889
   0.38889   0.38889   0.94444
   0.94444   0.27778   0.05556
   0.77778   0.11111   0.22222
   0.72222   0.38889   0.27778
   0.55556   0.22222   0.44444
   0.50000   0.50000   0.50000
   0.33333   0.33333   0.66667
   0.27778   0.61111   0.72222
   0.11111   0.44444   0.88889
   0.05556   0.72222   0.94444
   0.66667   0.33333   0.00000
   0.61111   0.61111   0.05556
   0.44444   0.44444   0.22222
   0.38889   0.72222   0.27778
   0.22222   0.55556   0.44444
   0.16667   0.83333   0.50000
   0.00000   0.66667   0.66667
   0.33333   0.66667   0.00000
   0.27778   0.94444   0.05556
   0.11111   0.77778   0.22222
   0.88889   0.22222   0.77778
   0.83333   0.50000   0.83333
   0.94444   0.61111   0.38889
   0.77778   0.44444   0.55556
   0.72222   0.72222   0.61111
   0.55556   0.55556   0.77778
   0.50000   0.83333   0.83333
   0.88889   0.55556   0.11111
   0.83333   0.83333   0.16667
   0.66667   0.66667   0.33333
   0.61111   0.94444   0.38889
   0.44444   0.77778   0.55556
   0.22222   0.88889   0.77778
   0.55556   0.88889   0.11111
   0.94444   0.94444   0.72222
   0.77778   0.77778   0.88889
   0.88889   0.88889   0.44444
\end{verbatim}

\noindent\makebox[\linewidth]{\rule{16.5cm}{0.4pt}}

\begin{verbatim}
Structure C; tI(2/3;0/1,3/2,3/2) shift = 0.00000   0.40000   0.35000
3.42624
1.00000  -3.00000  -2.44949
-3.00000   1.00000  -2.44949
1.00000   1.00000  -0.81650
O    B
64   64
Direct
   0.00000   0.10000   0.17500
   0.12500   0.22500   0.42500
   0.00000   0.10000   0.67500
   0.12500   0.22500   0.92500
   0.00000   0.35000   0.17500
   0.12500   0.47500   0.42500
   0.00000   0.35000   0.67500
   0.12500   0.47500   0.92500
   0.00000   0.60000   0.17500
   0.12500   0.72500   0.42500
   0.00000   0.60000   0.67500
   0.12500   0.72500   0.92500
   0.00000   0.85000   0.17500
   0.12500   0.97500   0.42500
   0.00000   0.85000   0.67500
   0.12500   0.97500   0.92500
   0.25000   0.10000   0.17500
   0.37500   0.22500   0.42500
   0.25000   0.10000   0.67500
   0.37500   0.22500   0.92500
   0.25000   0.35000   0.17500
   0.37500   0.47500   0.42500
   0.25000   0.35000   0.67500
   0.37500   0.47500   0.92500
   0.25000   0.60000   0.17500
   0.37500   0.72500   0.42500
   0.25000   0.60000   0.67500
   0.37500   0.72500   0.92500
   0.25000   0.85000   0.17500
   0.37500   0.97500   0.42500
   0.25000   0.85000   0.67500
   0.37500   0.97500   0.92500
   0.50000   0.10000   0.17500
   0.62500   0.22500   0.42500
   0.50000   0.10000   0.67500
   0.62500   0.22500   0.92500
   0.50000   0.35000   0.17500
   0.62500   0.47500   0.42500
   0.50000   0.35000   0.67500
   0.62500   0.47500   0.92500
   0.50000   0.60000   0.17500
   0.62500   0.72500   0.42500
   0.50000   0.60000   0.67500
   0.62500   0.72500   0.92500
   0.50000   0.85000   0.17500
   0.62500   0.97500   0.42500
   0.50000   0.85000   0.67500
   0.62500   0.97500   0.92500
   0.75000   0.10000   0.17500
   0.87500   0.22500   0.42500
   0.75000   0.10000   0.67500
   0.87500   0.22500   0.92500
   0.75000   0.35000   0.17500
   0.87500   0.47500   0.42500
   0.75000   0.35000   0.67500
   0.87500   0.47500   0.92500
   0.75000   0.60000   0.17500
   0.87500   0.72500   0.42500
   0.75000   0.60000   0.67500
   0.87500   0.72500   0.92500
   0.75000   0.85000   0.17500
   0.87500   0.97500   0.42500
   0.75000   0.85000   0.67500
   0.87500   0.97500   0.92500
   0.18750   0.93750   0.12500
   0.87500   0.87500   0.25000
   0.75000   0.75000   0.00000
   0.68750   0.93750   0.62500
   0.68750   0.93750   0.12500
   0.56250   0.81250   0.37500
   0.43750   0.68750   0.12500
   0.37500   0.87500   0.75000
   0.37500   0.87500   0.25000
   0.25000   0.75000   0.50000
   0.25000   0.75000   0.00000
   0.12500   0.62500   0.25000
   0.00000   0.50000   0.00000
   0.18750   0.93750   0.62500
   0.06250   0.81250   0.87500
   0.06250   0.81250   0.37500
   0.93750   0.68750   0.62500
   0.93750   0.68750   0.12500
   0.81250   0.56250   0.37500
   0.68750   0.43750   0.12500
   0.87500   0.87500   0.75000
   0.75000   0.75000   0.50000
   0.62500   0.62500   0.75000
   0.62500   0.62500   0.25000
   0.50000   0.50000   0.50000
   0.50000   0.50000   0.00000
   0.37500   0.37500   0.25000
   0.25000   0.25000   0.00000
   0.56250   0.81250   0.87500
   0.43750   0.68750   0.62500
   0.31250   0.56250   0.87500
   0.31250   0.56250   0.37500
   0.18750   0.43750   0.62500
   0.18750   0.43750   0.12500
   0.06250   0.31250   0.37500
   0.12500   0.62500   0.75000
   0.00000   0.50000   0.50000
   0.93750   0.18750   0.12500
   0.87500   0.37500   0.75000
   0.87500   0.37500   0.25000
   0.75000   0.25000   0.50000
   0.75000   0.25000   0.00000
   0.62500   0.12500   0.25000
   0.50000   0.00000   0.00000
   0.81250   0.56250   0.87500
   0.68750   0.43750   0.62500
   0.56250   0.31250   0.87500
   0.56250   0.31250   0.37500
   0.43750   0.18750   0.62500
   0.43750   0.18750   0.12500
   0.31250   0.06250   0.37500
   0.37500   0.37500   0.75000
   0.25000   0.25000   0.50000
   0.12500   0.12500   0.75000
   0.12500   0.12500   0.25000
   0.00000   0.00000   0.50000
   0.00000   0.00000   0.00000
   0.06250   0.31250   0.87500
   0.93750   0.18750   0.62500
   0.81250   0.06250   0.87500
   0.81250   0.06250   0.37500
   0.62500   0.12500   0.75000
   0.50000   0.00000   0.50000
   0.31250   0.06250   0.87500
\end{verbatim}

\noindent\makebox[\linewidth]{\rule{16.5cm}{0.4pt}}

\begin{verbatim}
Structure D; tF(1/2;-2/1,1/1,-1/1) shift = 0.00000   0.15000   0.50000
4.61084
-3.00000   5.00000   1.41421
-1.00000  -1.00000   1.41421
3.00000   1.00000   2.82843
O    B
288  288
Direct
   0.08333   0.07500   0.00000
   0.00000   0.32500   0.00000
   0.00000   0.07500   0.08333
   0.08333   0.32500   0.08333
   0.08333   0.07500   0.16667
   0.00000   0.32500   0.16667
   0.00000   0.07500   0.25000
   0.08333   0.32500   0.25000
   0.08333   0.07500   0.33333
   0.00000   0.32500   0.33333
   0.00000   0.07500   0.41667
   0.08333   0.32500   0.41667
   0.08333   0.07500   0.50000
   0.00000   0.32500   0.50000
   0.00000   0.07500   0.58333
   0.08333   0.32500   0.58333
   0.08333   0.07500   0.66667
   0.00000   0.32500   0.66667
   0.00000   0.07500   0.75000
   0.08333   0.32500   0.75000
   0.08333   0.07500   0.83333
   0.00000   0.32500   0.83333
   0.00000   0.07500   0.91667
   0.08333   0.32500   0.91667
   0.08333   0.57500   0.00000
   0.00000   0.82500   0.00000
   0.00000   0.57500   0.08333
   0.08333   0.82500   0.08333
   0.08333   0.57500   0.16667
   0.00000   0.82500   0.16667
   0.00000   0.57500   0.25000
   0.08333   0.82500   0.25000
   0.08333   0.57500   0.33333
   0.00000   0.82500   0.33333
   0.00000   0.57500   0.41667
   0.08333   0.82500   0.41667
   0.08333   0.57500   0.50000
   0.00000   0.82500   0.50000
   0.00000   0.57500   0.58333
   0.08333   0.82500   0.58333
   0.08333   0.57500   0.66667
   0.00000   0.82500   0.66667
   0.00000   0.57500   0.75000
   0.08333   0.82500   0.75000
   0.08333   0.57500   0.83333
   0.00000   0.82500   0.83333
   0.00000   0.57500   0.91667
   0.08333   0.82500   0.91667
   0.25000   0.07500   0.00000
   0.16667   0.32500   0.00000
   0.16667   0.07500   0.08333
   0.25000   0.32500   0.08333
   0.25000   0.07500   0.16667
   0.16667   0.32500   0.16667
   0.16667   0.07500   0.25000
   0.25000   0.32500   0.25000
   0.25000   0.07500   0.33333
   0.16667   0.32500   0.33333
   0.16667   0.07500   0.41667
   0.25000   0.32500   0.41667
   0.25000   0.07500   0.50000
   0.16667   0.32500   0.50000
   0.16667   0.07500   0.58333
   0.25000   0.32500   0.58333
   0.25000   0.07500   0.66667
   0.16667   0.32500   0.66667
   0.16667   0.07500   0.75000
   0.25000   0.32500   0.75000
   0.25000   0.07500   0.83333
   0.16667   0.32500   0.83333
   0.16667   0.07500   0.91667
   0.25000   0.32500   0.91667
   0.25000   0.57500   0.00000
   0.16667   0.82500   0.00000
   0.16667   0.57500   0.08333
   0.25000   0.82500   0.08333
   0.25000   0.57500   0.16667
   0.16667   0.82500   0.16667
   0.16667   0.57500   0.25000
   0.25000   0.82500   0.25000
   0.25000   0.57500   0.33333
   0.16667   0.82500   0.33333
   0.16667   0.57500   0.41667
   0.25000   0.82500   0.41667
   0.25000   0.57500   0.50000
   0.16667   0.82500   0.50000
   0.16667   0.57500   0.58333
   0.25000   0.82500   0.58333
   0.25000   0.57500   0.66667
   0.16667   0.82500   0.66667
   0.16667   0.57500   0.75000
   0.25000   0.82500   0.75000
   0.25000   0.57500   0.83333
   0.16667   0.82500   0.83333
   0.16667   0.57500   0.91667
   0.25000   0.82500   0.91667
   0.41667   0.07500   0.00000
   0.33333   0.32500   0.00000
   0.33333   0.07500   0.08333
   0.41667   0.32500   0.08333
   0.41667   0.07500   0.16667
   0.33333   0.32500   0.16667
   0.33333   0.07500   0.25000
   0.41667   0.32500   0.25000
   0.41667   0.07500   0.33333
   0.33333   0.32500   0.33333
   0.33333   0.07500   0.41667
   0.41667   0.32500   0.41667
   0.41667   0.07500   0.50000
   0.33333   0.32500   0.50000
   0.33333   0.07500   0.58333
   0.41667   0.32500   0.58333
   0.41667   0.07500   0.66667
   0.33333   0.32500   0.66667
   0.33333   0.07500   0.75000
   0.41667   0.32500   0.75000
   0.41667   0.07500   0.83333
   0.33333   0.32500   0.83333
   0.33333   0.07500   0.91667
   0.41667   0.32500   0.91667
   0.41667   0.57500   0.00000
   0.33333   0.82500   0.00000
   0.33333   0.57500   0.08333
   0.41667   0.82500   0.08333
   0.41667   0.57500   0.16667
   0.33333   0.82500   0.16667
   0.33333   0.57500   0.25000
   0.41667   0.82500   0.25000
   0.41667   0.57500   0.33333
   0.33333   0.82500   0.33333
   0.33333   0.57500   0.41667
   0.41667   0.82500   0.41667
   0.41667   0.57500   0.50000
   0.33333   0.82500   0.50000
   0.33333   0.57500   0.58333
   0.41667   0.82500   0.58333
   0.41667   0.57500   0.66667
   0.33333   0.82500   0.66667
   0.33333   0.57500   0.75000
   0.41667   0.82500   0.75000
   0.41667   0.57500   0.83333
   0.33333   0.82500   0.83333
   0.33333   0.57500   0.91667
   0.41667   0.82500   0.91667
   0.58333   0.07500   0.00000
   0.50000   0.32500   0.00000
   0.50000   0.07500   0.08333
   0.58333   0.32500   0.08333
   0.58333   0.07500   0.16667
   0.50000   0.32500   0.16667
   0.50000   0.07500   0.25000
   0.58333   0.32500   0.25000
   0.58333   0.07500   0.33333
   0.50000   0.32500   0.33333
   0.50000   0.07500   0.41667
   0.58333   0.32500   0.41667
   0.58333   0.07500   0.50000
   0.50000   0.32500   0.50000
   0.50000   0.07500   0.58333
   0.58333   0.32500   0.58333
   0.58333   0.07500   0.66667
   0.50000   0.32500   0.66667
   0.50000   0.07500   0.75000
   0.58333   0.32500   0.75000
   0.58333   0.07500   0.83333
   0.50000   0.32500   0.83333
   0.50000   0.07500   0.91667
   0.58333   0.32500   0.91667
   0.58333   0.57500   0.00000
   0.50000   0.82500   0.00000
   0.50000   0.57500   0.08333
   0.58333   0.82500   0.08333
   0.58333   0.57500   0.16667
   0.50000   0.82500   0.16667
   0.50000   0.57500   0.25000
   0.58333   0.82500   0.25000
   0.58333   0.57500   0.33333
   0.50000   0.82500   0.33333
   0.50000   0.57500   0.41667
   0.58333   0.82500   0.41667
   0.58333   0.57500   0.50000
   0.50000   0.82500   0.50000
   0.50000   0.57500   0.58333
   0.58333   0.82500   0.58333
   0.58333   0.57500   0.66667
   0.50000   0.82500   0.66667
   0.50000   0.57500   0.75000
   0.58333   0.82500   0.75000
   0.58333   0.57500   0.83333
   0.50000   0.82500   0.83333
   0.50000   0.57500   0.91667
   0.58333   0.82500   0.91667
   0.75000   0.07500   0.00000
   0.66667   0.32500   0.00000
   0.66667   0.07500   0.08333
   0.75000   0.32500   0.08333
   0.75000   0.07500   0.16667
   0.66667   0.32500   0.16667
   0.66667   0.07500   0.25000
   0.75000   0.32500   0.25000
   0.75000   0.07500   0.33333
   0.66667   0.32500   0.33333
   0.66667   0.07500   0.41667
   0.75000   0.32500   0.41667
   0.75000   0.07500   0.50000
   0.66667   0.32500   0.50000
   0.66667   0.07500   0.58333
   0.75000   0.32500   0.58333
   0.75000   0.07500   0.66667
   0.66667   0.32500   0.66667
   0.66667   0.07500   0.75000
   0.75000   0.32500   0.75000
   0.75000   0.07500   0.83333
   0.66667   0.32500   0.83333
   0.66667   0.07500   0.91667
   0.75000   0.32500   0.91667
   0.75000   0.57500   0.00000
   0.66667   0.82500   0.00000
   0.66667   0.57500   0.08333
   0.75000   0.82500   0.08333
   0.75000   0.57500   0.16667
   0.66667   0.82500   0.16667
   0.66667   0.57500   0.25000
   0.75000   0.82500   0.25000
   0.75000   0.57500   0.33333
   0.66667   0.82500   0.33333
   0.66667   0.57500   0.41667
   0.75000   0.82500   0.41667
   0.75000   0.57500   0.50000
   0.66667   0.82500   0.50000
   0.66667   0.57500   0.58333
   0.75000   0.82500   0.58333
   0.75000   0.57500   0.66667
   0.66667   0.82500   0.66667
   0.66667   0.57500   0.75000
   0.75000   0.82500   0.75000
   0.75000   0.57500   0.83333
   0.66667   0.82500   0.83333
   0.66667   0.57500   0.91667
   0.75000   0.82500   0.91667
   0.91667   0.07500   0.00000
   0.83333   0.32500   0.00000
   0.83333   0.07500   0.08333
   0.91667   0.32500   0.08333
   0.91667   0.07500   0.16667
   0.83333   0.32500   0.16667
   0.83333   0.07500   0.25000
   0.91667   0.32500   0.25000
   0.91667   0.07500   0.33333
   0.83333   0.32500   0.33333
   0.83333   0.07500   0.41667
   0.91667   0.32500   0.41667
   0.91667   0.07500   0.50000
   0.83333   0.32500   0.50000
   0.83333   0.07500   0.58333
   0.91667   0.32500   0.58333
   0.91667   0.07500   0.66667
   0.83333   0.32500   0.66667
   0.83333   0.07500   0.75000
   0.91667   0.32500   0.75000
   0.91667   0.07500   0.83333
   0.83333   0.32500   0.83333
   0.83333   0.07500   0.91667
   0.91667   0.32500   0.91667
   0.91667   0.57500   0.00000
   0.83333   0.82500   0.00000
   0.83333   0.57500   0.08333
   0.91667   0.82500   0.08333
   0.91667   0.57500   0.16667
   0.83333   0.82500   0.16667
   0.83333   0.57500   0.25000
   0.91667   0.82500   0.25000
   0.91667   0.57500   0.33333
   0.83333   0.82500   0.33333
   0.83333   0.57500   0.41667
   0.91667   0.82500   0.41667
   0.91667   0.57500   0.50000
   0.83333   0.82500   0.50000
   0.83333   0.57500   0.58333
   0.91667   0.82500   0.58333
   0.91667   0.57500   0.66667
   0.83333   0.82500   0.66667
   0.83333   0.57500   0.75000
   0.91667   0.82500   0.75000
   0.91667   0.57500   0.83333
   0.83333   0.82500   0.83333
   0.83333   0.57500   0.91667
   0.91667   0.82500   0.91667
   0.94444   0.75000   0.02778
   0.63889   0.75000   0.05556
   0.75000   0.25000   0.00000
   0.69444   0.50000   0.02778
   0.75000   0.75000   0.00000
   0.77778   0.50000   0.11111
   0.72222   0.75000   0.13889
   0.83333   0.75000   0.08333
   0.80556   0.75000   0.22222
   0.88889   0.00000   0.05556
   0.83333   0.25000   0.08333
   0.94444   0.25000   0.02778
   0.88889   0.50000   0.05556
   0.91667   0.25000   0.16667
   0.86111   0.50000   0.19444
   0.97222   0.50000   0.13889
   0.91667   0.75000   0.16667
   0.94444   0.50000   0.27778
   0.88889   0.75000   0.30556
   0.97222   0.75000   0.38889
   0.97222   0.00000   0.13889
   0.25000   0.75000   0.00000
   0.38889   0.50000   0.05556
   0.33333   0.75000   0.08333
   0.44444   0.75000   0.02778
   0.41667   0.75000   0.16667
   0.50000   0.00000   0.00000
   0.44444   0.25000   0.02778
   0.50000   0.50000   0.00000
   0.52778   0.25000   0.11111
   0.47222   0.50000   0.13889
   0.58333   0.50000   0.08333
   0.52778   0.75000   0.11111
   0.55556   0.50000   0.22222
   0.50000   0.75000   0.25000
   0.61111   0.75000   0.19444
   0.58333   0.75000   0.33333
   0.58333   0.00000   0.08333
   0.69444   0.00000   0.02778
   0.63889   0.25000   0.05556
   0.66667   0.00000   0.16667
   0.61111   0.25000   0.19444
   0.72222   0.25000   0.13889
   0.66667   0.50000   0.16667
   0.69444   0.25000   0.27778
   0.63889   0.50000   0.30556
   0.75000   0.50000   0.25000
   0.69444   0.75000   0.27778
   0.72222   0.50000   0.38889
   0.66667   0.75000   0.41667
   0.77778   0.75000   0.36111
   0.75000   0.75000   0.50000
   0.77778   0.00000   0.11111
   0.75000   0.00000   0.25000
   0.86111   0.00000   0.19444
   0.80556   0.25000   0.22222
   0.83333   0.00000   0.33333
   0.77778   0.25000   0.36111
   0.88889   0.25000   0.30556
   0.83333   0.50000   0.33333
   0.86111   0.25000   0.44444
   0.80556   0.50000   0.47222
   0.91667   0.50000   0.41667
   0.86111   0.75000   0.44444
   0.88889   0.50000   0.55556
   0.83333   0.75000   0.58333
   0.94444   0.75000   0.52778
   0.91667   0.75000   0.66667
   0.94444   0.00000   0.27778
   0.91667   0.00000   0.41667
   0.97222   0.25000   0.38889
   0.94444   0.25000   0.52778
   0.97222   0.50000   0.63889
   0.00000   0.50000   0.00000
   0.02778   0.75000   0.11111
   0.13889   0.25000   0.05556
   0.08333   0.50000   0.08333
   0.19444   0.50000   0.02778
   0.13889   0.75000   0.05556
   0.16667   0.50000   0.16667
   0.11111   0.75000   0.19444
   0.22222   0.75000   0.13889
   0.19444   0.75000   0.27778
   0.19444   0.00000   0.02778
   0.25000   0.25000   0.00000
   0.27778   0.00000   0.11111
   0.22222   0.25000   0.13889
   0.33333   0.25000   0.08333
   0.27778   0.50000   0.11111
   0.30556   0.25000   0.22222
   0.25000   0.50000   0.25000
   0.36111   0.50000   0.19444
   0.30556   0.75000   0.22222
   0.33333   0.50000   0.33333
   0.27778   0.75000   0.36111
   0.38889   0.75000   0.30556
   0.36111   0.75000   0.44444
   0.38889   0.00000   0.05556
   0.36111   0.00000   0.19444
   0.47222   0.00000   0.13889
   0.41667   0.25000   0.16667
   0.44444   0.00000   0.27778
   0.38889   0.25000   0.30556
   0.50000   0.25000   0.25000
   0.44444   0.50000   0.27778
   0.47222   0.25000   0.38889
   0.41667   0.50000   0.41667
   0.52778   0.50000   0.36111
   0.47222   0.75000   0.38889
   0.50000   0.50000   0.50000
   0.44444   0.75000   0.52778
   0.55556   0.75000   0.47222
   0.52778   0.75000   0.61111
   0.55556   0.00000   0.22222
   0.52778   0.00000   0.36111
   0.63889   0.00000   0.30556
   0.58333   0.25000   0.33333
   0.61111   0.00000   0.44444
   0.55556   0.25000   0.47222
   0.66667   0.25000   0.41667
   0.61111   0.50000   0.44444
   0.63889   0.25000   0.55556
   0.58333   0.50000   0.58333
   0.69444   0.50000   0.52778
   0.63889   0.75000   0.55556
   0.66667   0.50000   0.66667
   0.61111   0.75000   0.69444
   0.72222   0.75000   0.63889
   0.69444   0.75000   0.77778
   0.72222   0.00000   0.38889
   0.69444   0.00000   0.52778
   0.80556   0.00000   0.47222
   0.75000   0.25000   0.50000
   0.77778   0.00000   0.61111
   0.72222   0.25000   0.63889
   0.83333   0.25000   0.58333
   0.77778   0.50000   0.61111
   0.80556   0.25000   0.72222
   0.75000   0.50000   0.75000
   0.86111   0.50000   0.69444
   0.80556   0.75000   0.72222
   0.83333   0.50000   0.83333
   0.77778   0.75000   0.86111
   0.88889   0.75000   0.80556
   0.86111   0.75000   0.94444
   0.88889   0.00000   0.55556
   0.86111   0.00000   0.69444
   0.97222   0.00000   0.63889
   0.91667   0.25000   0.66667
   0.94444   0.00000   0.77778
   0.88889   0.25000   0.80556
   0.94444   0.50000   0.77778
   0.97222   0.25000   0.88889
   0.91667   0.50000   0.91667
   0.97222   0.75000   0.88889
   0.00000   0.75000   0.25000
   0.00000   0.00000   0.00000
   0.08333   0.00000   0.08333
   0.02778   0.25000   0.11111
   0.05556   0.00000   0.22222
   0.00000   0.25000   0.25000
   0.11111   0.25000   0.19444
   0.05556   0.50000   0.22222
   0.08333   0.25000   0.33333
   0.02778   0.50000   0.36111
   0.13889   0.50000   0.30556
   0.08333   0.75000   0.33333
   0.11111   0.50000   0.44444
   0.05556   0.75000   0.47222
   0.16667   0.75000   0.41667
   0.13889   0.75000   0.55556
   0.16667   0.00000   0.16667
   0.13889   0.00000   0.30556
   0.25000   0.00000   0.25000
   0.19444   0.25000   0.27778
   0.22222   0.00000   0.38889
   0.16667   0.25000   0.41667
   0.27778   0.25000   0.36111
   0.22222   0.50000   0.38889
   0.25000   0.25000   0.50000
   0.19444   0.50000   0.52778
   0.30556   0.50000   0.47222
   0.25000   0.75000   0.50000
   0.27778   0.50000   0.61111
   0.22222   0.75000   0.63889
   0.33333   0.75000   0.58333
   0.30556   0.75000   0.72222
   0.33333   0.00000   0.33333
   0.30556   0.00000   0.47222
   0.41667   0.00000   0.41667
   0.36111   0.25000   0.44444
   0.38889   0.00000   0.55556
   0.33333   0.25000   0.58333
   0.44444   0.25000   0.52778
   0.38889   0.50000   0.55556
   0.41667   0.25000   0.66667
   0.36111   0.50000   0.69444
   0.47222   0.50000   0.63889
   0.41667   0.75000   0.66667
   0.44444   0.50000   0.77778
   0.38889   0.75000   0.80556
   0.50000   0.75000   0.75000
   0.47222   0.75000   0.88889
   0.50000   0.00000   0.50000
   0.47222   0.00000   0.63889
   0.58333   0.00000   0.58333
   0.52778   0.25000   0.61111
   0.55556   0.00000   0.72222
   0.50000   0.25000   0.75000
   0.61111   0.25000   0.69444
   0.55556   0.50000   0.72222
   0.58333   0.25000   0.83333
   0.52778   0.50000   0.86111
   0.63889   0.50000   0.80556
   0.58333   0.75000   0.83333
   0.61111   0.50000   0.94444
   0.55556   0.75000   0.97222
   0.66667   0.75000   0.91667
   0.66667   0.00000   0.66667
   0.63889   0.00000   0.80556
   0.75000   0.00000   0.75000
   0.69444   0.25000   0.77778
   0.72222   0.00000   0.88889
   0.66667   0.25000   0.91667
   0.77778   0.25000   0.86111
   0.72222   0.50000   0.88889
   0.80556   0.50000   0.97222
   0.83333   0.00000   0.83333
   0.80556   0.00000   0.97222
   0.91667   0.00000   0.91667
   0.86111   0.25000   0.94444
   0.02778   0.00000   0.36111
   0.00000   0.00000   0.50000
   0.05556   0.25000   0.47222
   0.00000   0.50000   0.50000
   0.02778   0.25000   0.61111
   0.08333   0.50000   0.58333
   0.02778   0.75000   0.61111
   0.05556   0.50000   0.72222
   0.00000   0.75000   0.75000
   0.11111   0.75000   0.69444
   0.08333   0.75000   0.83333
   0.11111   0.00000   0.44444
   0.08333   0.00000   0.58333
   0.19444   0.00000   0.52778
   0.13889   0.25000   0.55556
   0.16667   0.00000   0.66667
   0.11111   0.25000   0.69444
   0.22222   0.25000   0.63889
   0.16667   0.50000   0.66667
   0.19444   0.25000   0.77778
   0.13889   0.50000   0.80556
   0.25000   0.50000   0.75000
   0.19444   0.75000   0.77778
   0.22222   0.50000   0.88889
   0.16667   0.75000   0.91667
   0.27778   0.75000   0.86111
   0.27778   0.00000   0.61111
   0.25000   0.00000   0.75000
   0.36111   0.00000   0.69444
   0.30556   0.25000   0.72222
   0.33333   0.00000   0.83333
   0.27778   0.25000   0.86111
   0.38889   0.25000   0.80556
   0.33333   0.50000   0.83333
   0.36111   0.25000   0.94444
   0.30556   0.50000   0.97222
   0.41667   0.50000   0.91667
   0.36111   0.75000   0.94444
   0.44444   0.00000   0.77778
   0.41667   0.00000   0.91667
   0.52778   0.00000   0.86111
   0.47222   0.25000   0.88889
   0.55556   0.25000   0.97222
   0.61111   0.00000   0.94444
   0.00000   0.25000   0.75000
   0.02778   0.50000   0.86111
   0.05556   0.75000   0.97222
   0.05556   0.00000   0.72222
   0.02778   0.00000   0.86111
   0.13889   0.00000   0.80556
   0.08333   0.25000   0.83333
   0.11111   0.00000   0.94444
   0.05556   0.25000   0.97222
   0.16667   0.25000   0.91667
   0.11111   0.50000   0.94444
   0.22222   0.00000   0.88889
   0.30556   0.00000   0.97222
\end{verbatim}

\end{document}